\documentclass[12pt, a4paper]{article}
\usepackage{scicite}
\newenvironment{sciabstract}{%
\begin{quote} \bf}
{\end{quote}}

\usepackage{amsmath, amssymb}
\usepackage{authblk}
\usepackage{graphicx}
\usepackage{cite}
\usepackage{hyperref}

\usepackage[margin=1in]{geometry}

\usepackage[super,sort&compress,comma]{natbib}

\usepackage{color}
\usepackage{cancel}

\usepackage{graphicx}
\usepackage{dcolumn}
\usepackage{bm}
\usepackage{siunitx} 
\usepackage{amsmath}
\usepackage{blkarray}
\usepackage[utf8]{inputenc}
\usepackage[T1]{fontenc}
\usepackage{amsthm}
\usepackage{scrextend}
\usepackage{subcaption}

\usepackage{soul}
\usepackage[normalem]{ulem}

\usepackage{algorithm2e}
\usepackage{tikz}

\DeclareMathOperator{\ArgMax}{ArgMax}
\DeclareMathOperator{\Roff}{R_{off}}
\DeclareMathOperator{\Ron}{R_{on}}
\newcommand{\comm}[1]{}

\title{Uncontrolled learning: co-design of neuromorphic hardware topology for neuromorphic algorithms}

\author[1,2]{F. Barrows}
\author[1,2]{J. Lin}
\author[1]{F. Caravelli}
\author[3,4]{D.R. Chialvo}

\affil[1]{Theoretical Division (T4), Los Alamos National Laboratory,\ Los Alamos, New Mexico 87545, USA}
\affil[2]{Center for Nonlinear Studies, Los Alamos National Laboratory, Los Alamos, New Mexico 87545, USA}
\affil[3]{Instituto de Ciencias F\'isicas (ICIFI-CONICET) Center for Complex Systems and Brain Sciences (CEMSC3), Escuela de Ciencia y Tecnolog\'ia, Universidad Nacional de Gral. San Mart\'in, Campus Miguelete, 25 de Mayo y Francia, 1650 San Mart\'in, Buenos Aires, Argentina}
\affil[4]{Consejo Nacional de Investigaciones Cient\'{\i}ficas y Tecnol\'ogicas (CONICET), Godoy Cruz 2290, 1425 Buenos Aires, Argentina}

\begin{document}

\baselineskip24pt

\maketitle

\begin{sciabstract}
Hardware-based neuromorphic computing remains an elusive goal with the potential to profoundly impact future technologies and deepen our understanding of emergent intelligence. The learning-from-mistakes algorithm is one of the few training algorithms inspired by the brain's simple learning rules, utilizing inhibition and pruning to demonstrate self-organized learning. Here we implement this algorithm in purely neuromorphic memristive hardware through a co-design process. 
This implementation requires evaluating hardware trade-offs and constraints. It has been shown that learning-from-mistakes successfully trains small networks to function as binary classifiers and perceptrons. However, without tailoring the hardware to the algorithm, performance decreases exponentially as the network size increases. When implementing neuromorphic algorithms on neuromorphic hardware, we investigate the trade-offs between depth, controllability, and capacity, the latter being the number of learnable patterns. 
We emphasize the significance of topology and the use of governing equations, demonstrating theoretical tools to aid in the co-design of neuromorphic hardware and algorithms. We provide quantitative techniques to evaluate the computational capacity of a neuromorphic device based on the measurements performed and the underlying circuit structure.
This approach shows that breaking the symmetry of a neural network can increase both the controllability and average network capacity. By pruning the circuit, neuromorphic algorithms in all-memristive device circuits leverage stochastic resources to drive local contrast in network weights. 
Our combined experimental and simulation efforts explore the parameters that make a network suited for displaying emergent intelligence from simple rules.
\end{sciabstract}

\section*{Introduction}

 The computational properties of the brain have motivated a virtuous cycle of innovation in computing which in turn has shaped our understanding of the brain. Technology influences the perception of the function of the brain; from Descartes hydraulic pump,\cite{Engelhardt2021-ic} to discussions between Turing and Jefferson about thinking machines, \cite{Turing_1950,Turing_1952}, through the cognitive revolution to the era of modern AI,\cite{Miller_2003,Fjelland_2020,George_2022,bubeck2023sparksartificialgeneralintelligence} our understanding of human intelligence has been profoundly influenced by our technology.  It is worth remembering Ada Lovelace cautioned that it was "desirable to guard against the possibility of exaggerated ideas that arise as to the powers" of machines.\cite{Menabrea_2015} Thus through judicious pairing of neuromorphic algorithms and hardware it seems we should continue this virtuous cycle.

 Neuromorphic algorithms, when implemented on neuromorphic hardware, hold promise as the foundation for next-generation computing architectures. By emulating the brain’s non-von Neumann architecture and utilizing local learning rules, these systems aim to replicate the brain’s emergent intelligence and complexity.\cite{Schuman_2022,Christensen_2022,Kim_Front_2024} One such algorithm is learning-from-mistakes, which remains one of the few biologically plausible methods designed for training neural networks.\cite{Chialvo_1999} Despite being discussed for over 25 years, a neuromorphic and bio-inspired variation of learning-from-mistakes has yet to be fully implemented in all-neuromorphic hardware\cite{Bak_2001,Wakeling_2003,Brigham_2009}, although the original Bak-Chialvo with memristive devices has been recently studied. \cite{Carbajal_2022,Nikiruy_2024}
In part, this is because there are numerous hardware challenges we need to overcome through software and hardware co-design. In effect, we need to rig the "hardware lottery" and tailor hardware and algorithms for each other.\cite{Hooker_2021} In the present manuscript, we will study both in hardware and theoretically the implementation of learning-from-mistakes with memristive devices, e.g. Ohmic components with memory whose resistance can be controlled via an external voltage

Learning-from-mistakes implements synaptic pruning, a process observed in the brain involving the removal of neuronal connections through inhibitory signals.\cite{MONTAGU_1964,CHECHIK_1999,Faust_2021} This pruning mechanism reduces synaptic connectivity, mirroring the brain’s method of refining neural networks.
Importantly learning-from-mistakes does not rely on classical reinforcement which would necessitate of an oracle. Instead, it is a model of continuous Hebbian learning wherein memristive devices are updated locally based on local electrical potential.\cite{Sanger_1989,Chialvo_1999,Carbajal_2022} Only corrective signals are given to the network when there is an incorrect association. Corrections propagate in the forward direction via an inhibitory signal. 
This approach aligns with the brain’s method of learning from errors without explicit supervision or reinforcement.
Similarly, the brain learns explicit supervision or reinforcement
and must learn what types of activity minimize errors.\cite{Sanger_1989} Learning-from-mistakes is neither unsupervised nor supervised training, instead it attempts to minimize error against an external reference, e.g., minimize surprise.\cite{Friston_2010}

In contrast to other training algorithms, including backpropagation and equilibrium propagation, learning-from-mistakes displays learning from functions that could be implemented in the brain that are both self-organized and display phase transitions.\cite{Wakeling_2003} Backpropagation requires signals to propagate backward through the network,\cite{rosenblatt1961principles,Linnainmaa_1976} while equilibrium propagation requires free and clamped phases.\cite{Scellier_2017} This is important as all-neuromorphic hardware often lacks the complexity of traditional computing systems.


The perceptron is a mathematical model of a biological neuron, 
learning-from-mistakes was designed for perceptron-based networks.\cite{Block_1962,rosenblatt1957perceptron} In this design only a single neuron in any hidden layer is active at a time. This process is not immediately amenable to neuromorphic hardware as it necessitates additional CMOS complexity. Neuromorphic materials may be simpler in functionality than a perceptron, lacking integrated logic despite possessing memory. 

Previous work to build perceptrons from small functional units of memristive devices required more complicated training algorithms, e.g., backpropagation, and did not examine the scaling in the size of the networks to larger pattern recognition.\cite{Silva_2020} 
Simple neuromorphic computing platforms are physically realizable but understanding how these architectures scale and how to incorporate neuromorphic algorithms has been little explored.

More carefully phrased, memristive devices are neuromorphic circuit elements with variable resistance, the resistance depends on the history of applied bias or current and thus functions as a memory.\cite{Strukov_2008,reviewCarCar,Kumar_2022,Xiao_2023} Changing resistance is analogous to changing synaptic weights in a neural network.
Understanding the properties and emergent logic of neuromorphic networks is an essential step in developing neuromorphic computation platforms. The physical properties of memristive devices can be leveraged for computing, including volatility and stochasticity.\cite{Mambretti2022,Gaba_2013,Ignatov_2017,Lin_2024} This contrasts with previous work wherein CMOS control of a memristive circuit is used to introduce non-biological logic operations within the hidden layer of the network.\cite{Nikiruy_2024}

In this work, we present a modern approach to Bak-Chialvo learning-from-mistakes training algorithm and demonstrate how to implement neuromorphic algorithms on neuromorphic hardware. 
In particular, we demonstrate its implementation on a purely memristive circuit, i.e. in the absence of active components and passive elements other than memristive devices.  Moreover, we provide the first experimental demonstration of neuromorphic training on an all-neuromorphic material, emphasizing an approach that eschews control of individual circuit elements.
 
There is currently no unified framework for tailoring circuits to implement neuromorphic algorithms effectively. This work addresses several challenges, including evaluating network capacity, controllability, and correlations. The capacity is the total number of learnable patterns. One comment is that learnable and trainable patterns are two different notions. We refer to ``learnable patterns" as the number of patterns that, if one were able to change each resistance, the pattern would be learned. A ``trainable pattern" instead refers to the fact that the pattern can be reached via the control, e.g. the network is controllable.
We outline methods for addressing tradeoffs between these circuit properties. We identify methods to enhance algorithm performance by hardware design and capacity while maintaining control.

 The work presented here is generalizable to complex networks with nonlinear dynamics and extensive connections that cannot be independently controlled. 

\section*{Results}
\subsection*{Learning-from-mistakes}

The memristive networks we will study in this manuscript can be decomposed into three parts. We identify a set of bulk nodes (composed of one or multiple hidden layers), and inputs and output nodes. Input nodes are connected to a voltage generator, while at the output nodes, we can read currents via an ammeter.
Given this setup, the learning-from-mistakes algorithm trains a set of input-output mappings, collectively referred to as a pattern. In our implementation, learning-from-mistakes train patterns without accessing or directly controlling the memristors within the network, instead applying bias and performing measurements solely on the network inputs and outputs, respectively. Each mapping specifies an input node where a positive bias is applied and an output node where current is desired. For example, given a mapping $\left[
    1,2
\right]$, a bias is applied to the input node indexed at 1, and the network's output should be registered at the output node indexed at 2. This is a binary classifier, which has already been considered in the neuromorphic literature (see for instance \cite{Mirigliano2021}).

More precisely, the learning-from-mistakes algorithm is detailed in Algorithm \ref{alg:LearnFromMistakes}:
\begin{algorithm}[h!]
\caption{Pseudocode implementation of learning-from-mistakes in {memristive} networks.}\label{alg:cap}
\KwData{$M$: set of mappings, $\vec{v}$: applied bias, $\vec{i}$: current measured}
\While{ $\vert \text{Correct}\vert < \vert \text{M}\vert$}{
$m \gets \text{Get Random} (M)$\\
$\textbf{Apply: } v_\text{read} \rightarrow \left[ m_\text{input} , \text{All Outputs}\right]$\\
$\textbf{Measure: } \vec{i}$\\
\eIf{ $\text{ArgMax}(\vec{i})=m_\text{output}$}{
$\text{Correct}\gets m$}{ 
$\textbf{Delete: }(\text{Correct},m)$\\
$\textbf{Apply: } v_\text{correct} \rightarrow\left[ m_\text{input} , \text{ArgMax}(\vec{i})\right]$\\
$\textbf{Apply: } v_\text{normalize} \rightarrow \left[ \text{All Inputs} , \text{All Outputs}\right] $\\
}
}
\label{alg:LearnFromMistakes}
\end{algorithm}

Training involves randomly selecting a mapping, applying a low bias $v_\text{read}$ to the input node, and measuring the current on all output nodes. The current measurements determine whether the network produces the correct output for a given mapping, assessing if the network recognizes the pattern. Current is measured on all output nodes, the output node with the maximum current is identified as the network's output, using an $\ArgMax$ function.

 If the measured output node matches the desired output node then no corrective step is needed.
If the output node is incorrect, then a correction is applied to the circuit. The circuit is altered such that all but the erroneous output {nodes} are disconnected from {the} ground, and a negative bias, $v_\text{correct}$, is applied to {decrease} the conductivity between the input and incorrect output node.

To prevent the {memristive devices} from being driven to a uniform high resistance state, after the correction, all output nodes are connected to {the} ground again, and a positive bias, $v_\text{normalize}$ is applied to all input nodes in an attempt to increase the conductivity of the network.

Training continues by randomly selecting another mapping for a read operation. This cycle repeats until all the mappings in a pattern have been trained and read correctly or training exceeds {a given} time limit. 
For any given circuit, the correction voltage is constant for every applied correction, it does not vary throughout {the }training, and is determined by {the} network size. 
At no point is the measured current used to change the correction strength.

The loss function, derived in the Supplementary Material, is a cross-entropy loss function,
\begin{align}
    CE^a_{j}=- p^a_j\ln(\delta^a_{j,g})
\end{align}
where $\delta^a_{j,g}$ is a delta function indicating whether the measured output $j$ of a mapping $a$ matches the desired output $g$. $p^a_j$ represents the probability the output node $j$ is the measured network output, as implemented $p^a_j$ is one or zero if the node $j$ does or does not have the maximum current. The cost function is binary; there is a uniform error signal whenever the $\ArgMax$ output is not on the desired output and no error when the $\ArgMax$ output is correct. 
The entropy is zero only when the output for the mappings aligns with the desired output.

  \subsection*{Training}
  	
\begin{figure}[t!] \centering
\includegraphics[width=.85\textwidth]{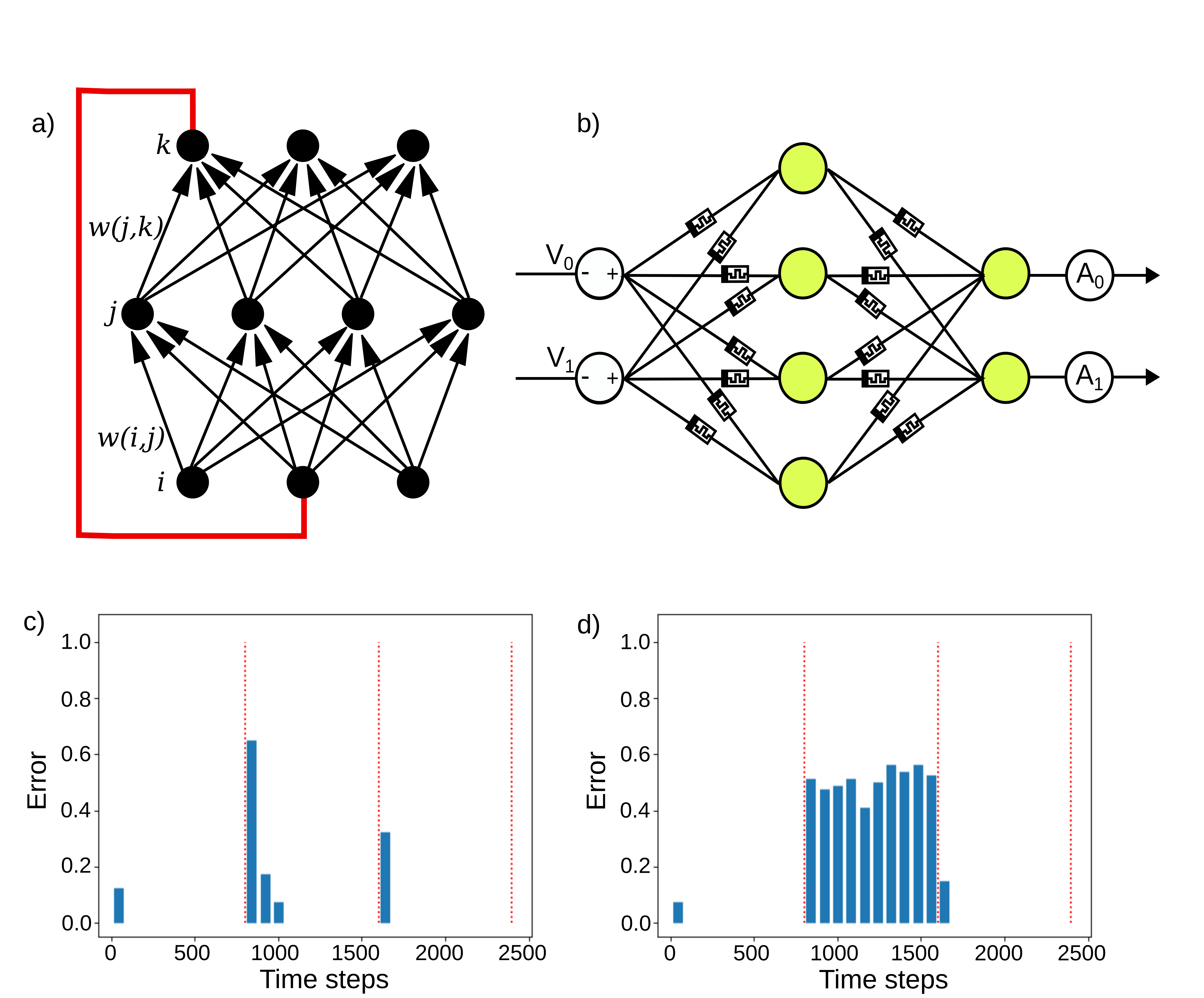}
\caption{ (a) Schematic of a network, network weights are indicated with \textit{w(i,j)}, correction signals are applied between an undesired output edge and the input edges, shown in red. (b) Circuit schematic of {a memristive} network. Bias is applied at the input nodes using a voltage generator, output currents are measured with ammeter $A_0$ and $A_1$. (c)  and (d) Error during training for two different $2$-input, 4 nodes in the hidden layer, and $2$-output {memristive} circuits. Two patterns are trained over three training eras, e.g., Pattern 1, Pattern 2, Pattern 1. Red dashed lines {indicate} the end of a training era and the switching of patterns. The error rate drops to zero when the pattern is learned. Red dashed lines indicate the end of a training session and switching the pattern which is being trained. (d) The network never learns the second pattern and thus the error rate remains high, The network quickly relearns the first pattern during the third training era. }
\label{fig:KNOWM_Exp}
\end{figure}

Experiments were conducted using networks of voltage-{controlled memristive devices}. Each network comprised $2$ input nodes, $2$ output nodes, and one hidden layer with $4$ nodes. Thus, each network consisted of $16$ {memristive component}. 
Layers are fully connected as illustrated in Figure \ref{fig:KNOWM_Exp}. Each connection between nodes consists of a {memristive device} hardwired to the nodes. Electrical bias could be applied independently to the input nodes, and current could be measured in parallel at the output nodes using ammeters. Experimental details are described in the Methods section.
  
  Training consisted of sequentially training two incompatible patterns, $\big\lbrace \left[ 0,1  \right]
,\left[1,0\right] \big\rbrace$ and $\big\lbrace \left[ 0,0\right], \left[1,1\right] \big\rbrace$. 
  These two patterns were alternated over three training sessions, with the previously trained state serving as the initial state for the subsequent training.
An epoch is defined as $80$ random mappings trained on the network. {Each training} pattern consists of $10$ epochs, as training consists of multiple epochs we call these a training era. {The training} was performed on fully connected neural networks with one hidden layer, as shown in Figure \ref{fig:KNOWM_Exp}.

{Some representative} results of training are shown in Figure \ref{fig:KNOWM_Exp} (c) and (d). Errors for each epoch are plotted, with red dashed lines indicating the end of training eras and changes in the pattern being trained. Networks can learn a pattern such that the error drops to zero, and the read bias and {memristive device} volatility {are} low such that the learned state is generally retained. A single network can learn multiple incompatible patterns, as evidenced in (c). In (d), the second pattern is not successfully learned, and the error does not drop to zero, during the third training era, the network quickly relearns the first pattern. These results are representative, with average errors from multiple training sessions provided in the supplementary material.
  
{The simulations} were performed to evaluate the behavior of individual {memristor} throughout training. {In particular,  for the numerical integration we} used the PySpice package\cite{PySpice} and the operator formalism {previously developed} \cite{caravelli2017complex,reviewCarCar}. Details of the simulations are provided in the Supplementary Material, along with experimental and simulated {IV curves}.

\begin{figure}[t!] \centering
\includegraphics[width=.9\textwidth]{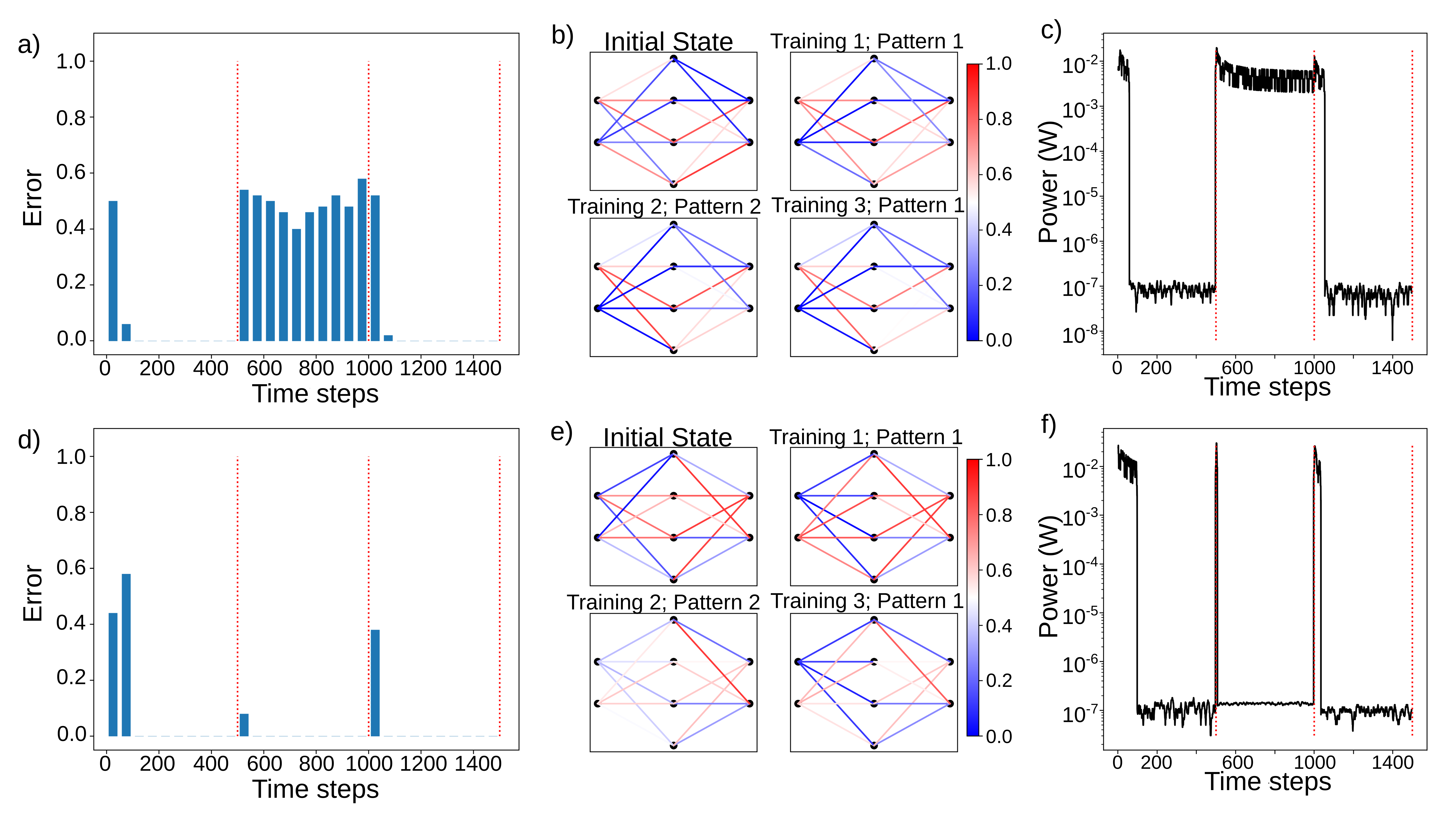}
\caption{(a) Error during training of a $2\times 4\times 2$ {memristive} network on two alternating patterns over three training sets. The network does not learn the second pattern but quickly relearns the first pattern during the third training era. (b) Initial and trained values of the memory parameter, $x$, in the network, $x=1$ is the conductive state with {a} resistance of $\Ron$, shown in red. (c) Power consumption during training protocol. Error (d),  memory parameter values (e), and power consumption during training of a second $2\times 4\times 2$ network.}
    \label{fig:KNOWM_spice}
\end{figure}

Networks with $2$ input nodes, $2$ output nodes and one hidden layer of $4$ nodes were simulated. Two alternating patterns are trained over three training eras. 
 Figure \ref{fig:KNOWM_spice} shows results for two different networks.


In Figure \ref{fig:KNOWM_spice} (a) and (d) the error during training is shown. The network is generally able to learn at least one pattern, after learning the first pattern the network is generally able to relearn this pattern rapidly in a later training era. 
Figure \ref{fig:KNOWM_spice} (b) and (e) {show schematics} of the {memristive} network at the end of training each pattern, with color indicating the resistance between $\Ron$ and $\Roff$, with $x=1$ {corresponding to} conductive state with resistance $\Ron$. 
The resistance in the {memristive devices} connecting the input and hidden layers diverges towards $\Ron$ or $\Roff$ values throughout training. {The memristive components} originating from a single input node often achieve a near-uniform high or low conductivity.
{The contrast} in resistance is also {apparent} on the second layer, where the two {memristive components} connecting from a single hidden layer node to the output layer frequently take contrasting high and low resistance values. 

{Instead, figure} \ref{fig:KNOWM_spice} (c) and (f) display the power consumption for operating the device during training. Training is energy intensive due to the read and the write operations. Notably, the power consumed during training for a specific pattern decreases over time, suggesting that the network minimizes power dissipation even if it settles into a stable but incorrect state, as observed in (c).

\subsubsection*{The capacity of larger networks}
{In this section we study the capacity of networks. We use the term capacity as (any) measure that provides an estimate of the number of patterns that can be learned.}

Larger {memristive} networks {can be} studied using the projection operator dynamical equation {derived in \cite{caravelli2017complex,Caravellisciad}}, 
\begin{equation}
    \dot{\vec{x}}=\frac{-1}{\beta}\left(I-\chi\Omega_A X\right)^{-1}\Omega_A\vec{v}_\text{source}-\alpha\vec{x},
\label{eqn:CaravelliEqn}
\end{equation}
the resistance state is parameterized with a memory parameter, $x$, which takes values $0\leq x\leq1$. {The matrix $X$ is diagonal, with $X_{ii}=x_i$.} For example, the resistance can be written $R(x)=\Ron x +(1-x)\Roff$. The dynamics of $x$ is driven by the external bias, $v_\text{source}$. Here $\beta$ is the learning rate, $\alpha$ is the volatility of the {device}, $\chi$ is a scaling factor defined as $\chi=\frac{\Roff-\Ron}{\Roff}$, and $\Omega_A$ is the loop projection operator that projects into the loop subspace of the circuit.\cite{barrows_2024} Simulations are described in the supplementary material.

For weakly volatile {devices}, $\alpha$ is small and can be neglected during rapid successive training.
The update function during training is $\dot{\vec{x}}$. When $\beta$ is uniform and considered as a scalar, the update function $\dot{x}$ is an eigenfunction of $\Omega_A$, 
\begin{equation}
    \Omega_A\dot{x}=\dot{x} .
    \label{eqn:LoopEigenfunction}
\end{equation}
 Thus, the network updates occur along the loops of the network, {and} parallel paths that form loops are similarly affected for any given update. As the size of the {memristive} network {increases,} the number of parallel paths also increases. Previous work examined the role of geometry in learning-from-mistakes but has not identified the critical role of topology in the geometries investigated.\cite{Wakeling_2003}

{Let us provide a few hints of the role that the network topology has in the learning architecture. First, an}
 optimal topology {must} have pathways between all input and output nodes, {each} individually controllable while retaining integration capacity through a connected hidden layer.
The fully connected layered network previously investigated with the learning-from-mistakes protocol\cite{Chialvo_1999,Carbajal_2022,Nikiruy_2024}, shown in Figure \ref{fig:KNOWM_Exp}, contains the maximum number of parallel cycles in a two-layered network. Other explored 
 networks include lattice topologies and random networks.\cite{Bak_2001}

\begin{figure}[h!] \centering
    \includegraphics[width=.7\textwidth]{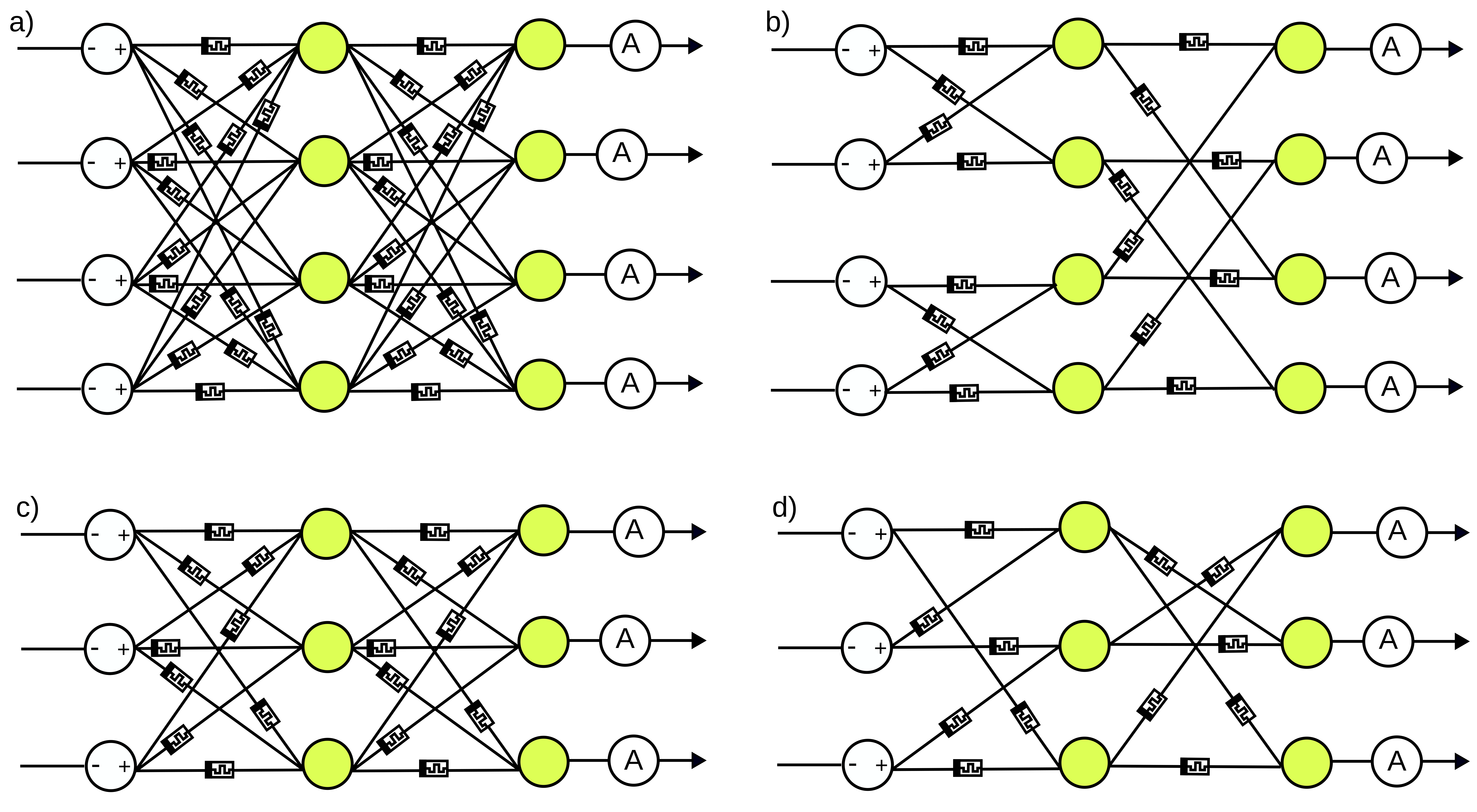}
    \caption{(a) and (b) Schematic of $4$-input and $4$-output network with $4$ nodes in the hidden layer, with fully connected and pruned topologies, respectively. 
    (c) and (d) Schematic of $3$-input and $3$-output network with $3$ nodes in the hidden layer, with fully connected and pruned topologies, respectively. }
    \label{fig:4x4Networks}
    \end{figure}

The role of network size and topology can be investigated through the network's capacity. By calculating the capacity we can compare {two different network structures}. Here, we focus {on} simple pruned networks to reduce the number of cycles, shown in Figure \ref{fig:4x4Networks} (b) and (d){. However,} the techniques developed {in the present manuscript} can be applied to any underlying network.
    

 Larger networks should be able to learn more mappings, corresponding to a larger capacity. As neuromorphic algorithms sacrifice some amount of {controllability (i.e. the ability to lead a single device to a particular resistive value via voltage)} to implement simple learning rules, errors, and crosstalk within the network will increase in larger networks. {This suggests the existence of} a trade-off between the capacity and controllability {when there is a bulk of devices that are not directly accessible}. Analytical tools to assess this trade-off enable co-design of hardware. Defining analytical tools to quantify controllability in nonlinear connected networks {is a nontrivial task. We then} make progress by finding closed-form expressions for {the} capacity. 
 It is important to distinguish between capacity, the maximum number of mappings a network can learn simultaneously, and the ability to achieve a state of maximum capacity with a specific training protocol. A network may have the potential for high capacity {states}, but it might still be challenging to configure {it} using a particular algorithm. 
When the {network's output} is the maximum current on the output nodes, {e.g.} a winner-take-all-algorithm, we can calculate the capacity using the $\ArgMax$ function. This can be written explicitly as
\begin{subequations}
\begin{align}
    \ArgMax^{s_k}(i_i \vert \vec{i}) &= \lim_{\frac{1}{k_B T}\rightarrow \infty} \frac{\exp(\frac{1}{k_B T} i_i)}{\sum_o \exp(\frac{1}{k_B T} i_o)}
    \\
    &= \lim_{\frac{1}{k_B T}\rightarrow \infty} \frac{\exp(\frac{1}{k_B T}\mathcal{P}_{ij}\Omega_{jk} s_k)}{\sum_o \exp(\frac{1}{k_B T}\mathcal{P}_{oj}\Omega_{jk} s_k)} ,
\end{align}
\label{eqn:ArgMaxProjection}
\end{subequations}
which is derived in the {Supplementary Material}.

 {In the capacity estimation above, we defined} {the state-dependent operator} $\mathcal{P}\equiv (I-\chi \Omega X)^{-1}$;  {it is interesting to observe that} $\mathcal{P}\Omega\vec{s}$ corresponds to the positive valued currents measured at the output nodes. Note that for any {resistive pattern, which here corresponds to a state $x$} there is only one nonzero value in $\vec{s}$, {and} each mapping can be characterized by {the biased input node}. For example, $s^k$ is a specific mapping with {a} bias applied to input node $k$. 
  We define an $OUT \times IN$  matrix $A$, {where the} columns correspond to the $IN$-mappings in a pattern, {while} the rows correspond to the $OUT$-output currents, {i.e.}
\begin{align}
    A=\begin{pmatrix}
        \ArgMax^{s_0}(i_0\vert\vec{i}), & \dots, & \ArgMax^{s_{IN}}(i_0\vert\vec{i})
        \\ \vdots & \ddots & \vdots 
        \\ \ArgMax^{s_0}(i_{OUT}\vert\vec{i}), & \dots, & \ArgMax^{s_{IN}}(i_{OUT}\vert\vec{i}) .
    \end{pmatrix}
    \label{eqn:StatMechCapacity}
\end{align}
As the output current at each node in a trained network is unique, $A$  takes values of $0$ and $1$, and each column will have one nonzero value. 
The rank of this matrix is the network's capacity, {which is,} the number of rules with distinct outputs. If each row has a distinct output, the rank of this matrix is $\text{min}(IN,OUT)$. In the {Supplementary Material} we prove that for a circuit with distinct outputs for all distinct inputs, the rank will be the minimum of the number of {input and output} edges.

\begin{figure}[t!] \centering
    \includegraphics[width=.9\textwidth]{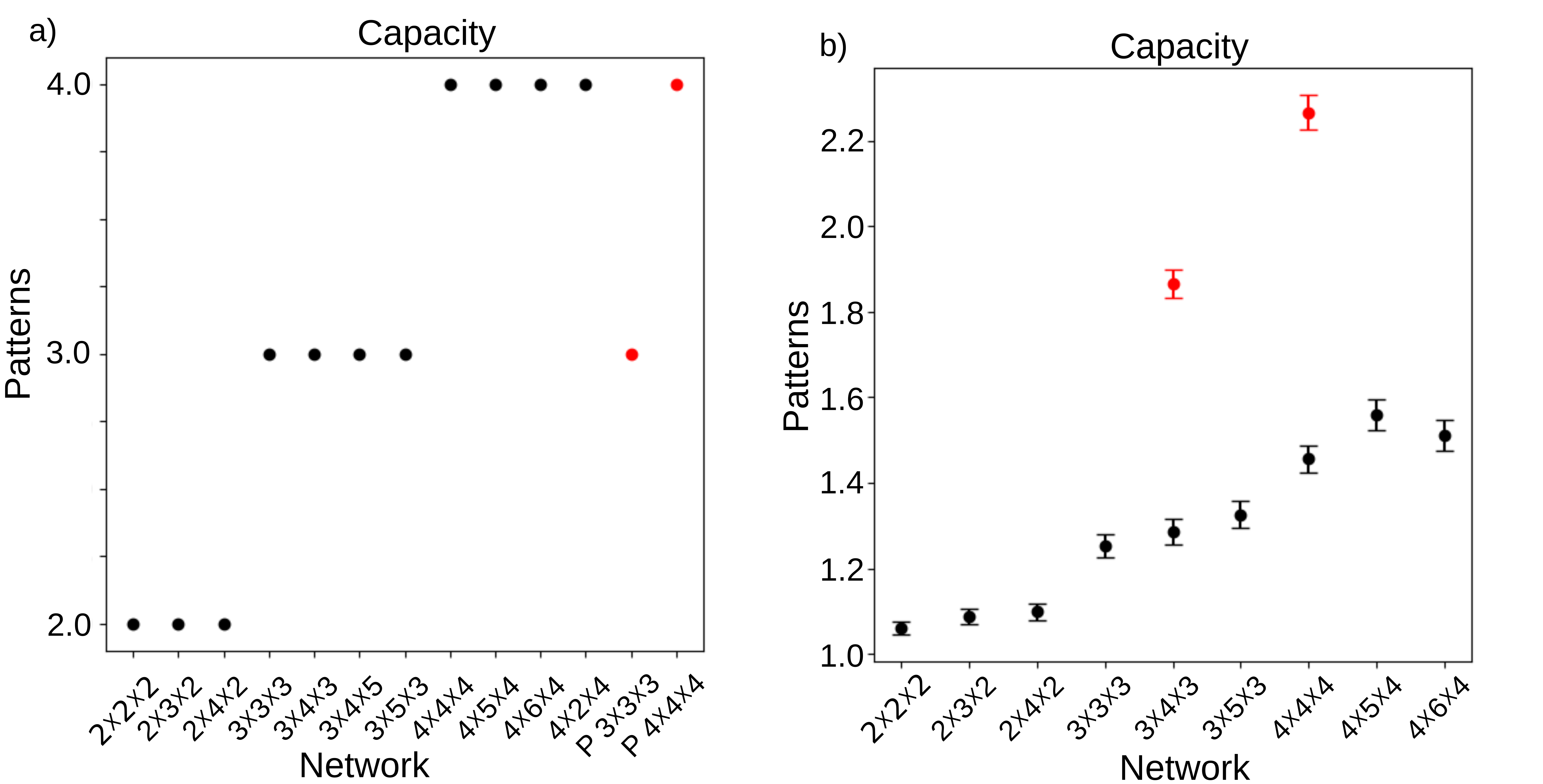}
         \caption{(a) Maximum capacity for different networks. (b) Ensemble capacity with $95\%$ confidence interval shown. {The red} dots represent the pruned networks. {We observe that pruned networks have higher capacity.}}
             \label{fig:Capacities}
\end{figure}

The maximum capacity of a winner-take-all algorithm for different circuits is shown in Figure \ref{fig:Capacities}, {where} pruned networks are labeled with the prefix $P$. The maximum capacity was determined as the maximum rank of randomly initialized circuits using equation \ref{eqn:StatMechCapacity}. The maximum capacity corresponds to the minimum of the number of input and output nodes when the network does not have a {bottleneck}. For fully connected two-layer networks, the number of nodes in the middle layer must be greater than $1$ to avoid a bottleneck, as shown in the {Supplementary Material}.

The average capacity of randomly initialized resistance values on these networks. 
This ensemble capacity increases with network size. For the fully connected two-layered networks, the ensemble capacity grows sublinearly with the number of input/output nodes. 
{We thus observe that pruned} networks exhibit a significantly larger ensemble capacity compared to fully connected networks. 

\subsubsection*{Learning contrasting weights}
\begin{figure}[t!] \centering
    \includegraphics[width=1\textwidth]{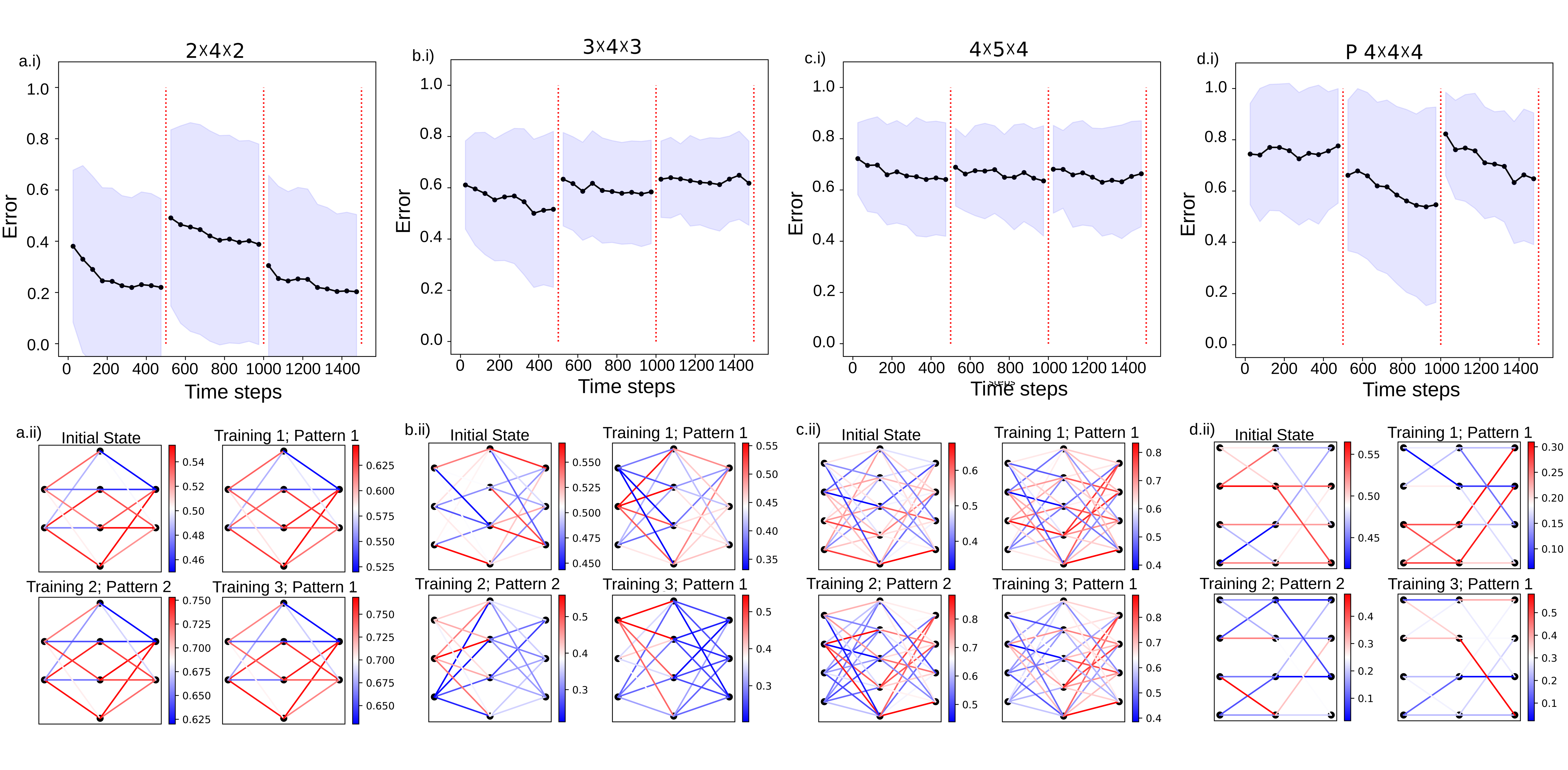}
     \caption{ Top row (.i): Average error of twenty different networks training two alternating patterns over three training eras, shown by the black line. The standard deviation is shown with the shaded area. Bottom row (.ii): Average values of the memory parameter $x$ at the beginning of training and at the end of each training era. }
     \label{fig:IntegrationCNN}   
\end{figure}

{We now discuss the details regarding the training.}
{The training of the memristive networks} is implemented on larger {circuits} as described above{. The }results are shown in Figure \ref{fig:IntegrationCNN}. 
The average error during training is depicted by the black line, {where the standard deviation is shown in the shaded region for networks of different sizes}. The average error {is consistent with}  experimental results, as discussed {in depth in Supplementary Material}.

{We observe that as} the number of input and output edges increases, the frequency at which the network learns a pattern decreases. With larger network sizes, the ability to train a given pattern diminishes, and the variation in the network error rate decreases.
Comparing the pruned network in column (d), we observe that the error is reduced compared to the $4$-in $4$-out fully connected two-layered network in column (c). The average network resistance of networks that successfully learn a pattern at the end of each training era is shown in the bottom row in Figure \ref{fig:IntegrationCNN}. 
When a pattern is successfully learned, a noticeable contrast in the resistance of {each memristive device} from nodes in the hidden layer is evident. This can be seen in the output layer of Figure \ref{fig:IntegrationCNN}. For example, in the second layer of the $4\times 5 \times 4$ network, at the end of training pattern 1 and pattern 2 there is a single blue edge from nodes in the hidden layer connected to the output layer. This contrast corresponds to a learned state.  Training this low resistance edge to connect to the correct output node becomes increasingly difficult as the network size grows, due to correlations in the effect of the training, as discussed below.
The resistive contrast in the output layer of the pruned $4 \times 4 \times 4$ network is comparable to the $2$-input $2$-output networks, where the contrast is between two {memristive devices}. Variation in the network's parameters is a resource that enables training and drives this contrast, as discussed in the supplementary material.

\begin{figure}[h!] \centering
    \centering
     \includegraphics[width=.7\textwidth]{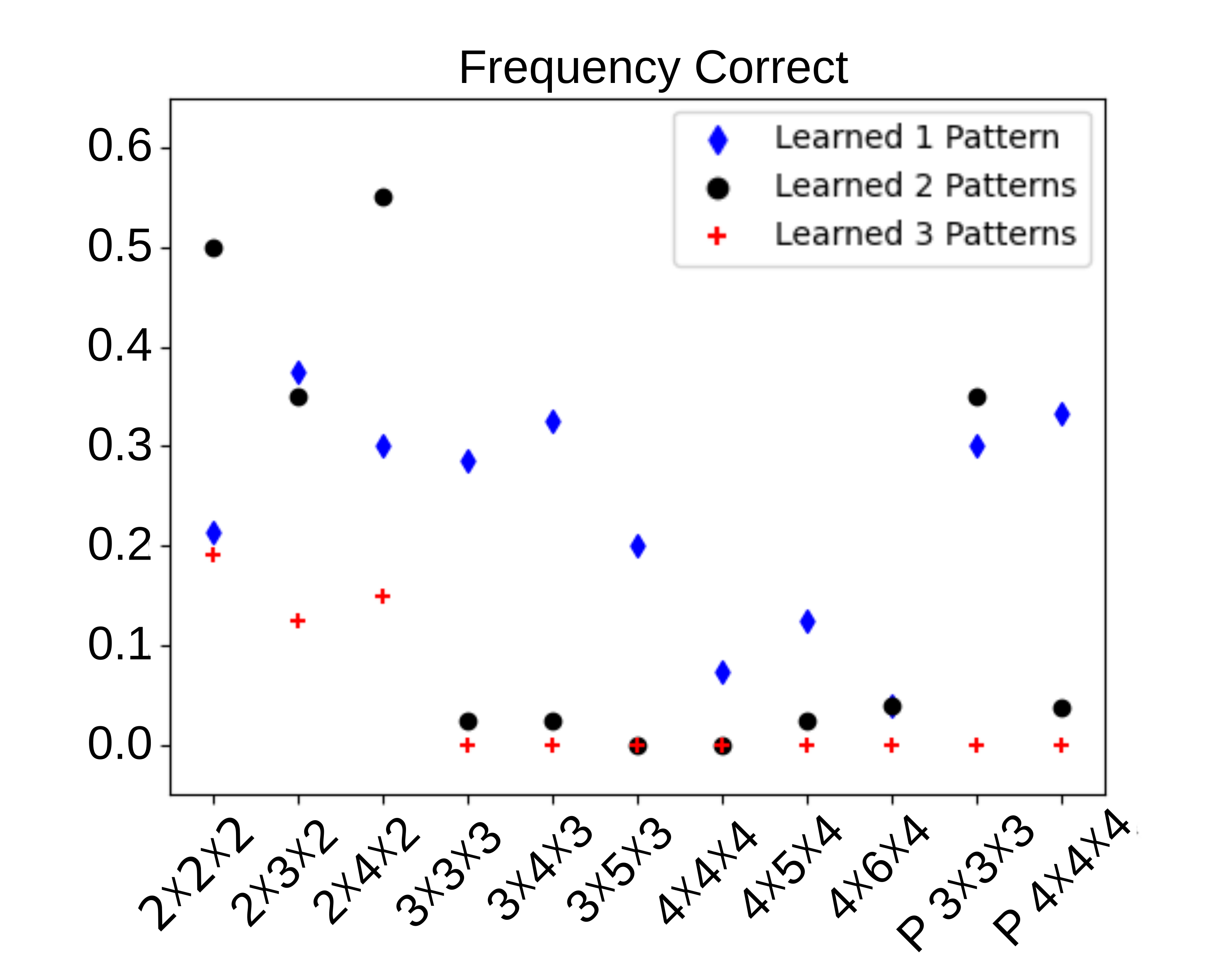}
            \caption{Percent of training sessions in which $1$, $2$, and $3$ patterns are successfully trained are shown by diamond \textcolor{blue}{$\blacklozenge$}, circle $\bullet$, and cross \textcolor{red}{$+$}, respectively}
\label{fig:AllError}
\end{figure}

The frequency that one, two, and three patterns are successfully learned during a training session is shown in Figure \ref{fig:AllError}. It is apparent that the $2$-in $2$-out networks learn at least one pattern in over 90\% of training sessions. 
{In contrast, we found that networks} with more than $2$ input and $2$ output nodes rarely learn more than one pattern per training session.

{This is one of the key differences between learnable and unlearnable states.} Trainability determines how often training succeeds, 
{and} is strongly determined by the number of input and output nodes, which also determines the number of mappings in a pattern. 
Notably, the modified pruned topology, $P 4\times 4\times 4$ and $P 3\times 3 \times 3$, exhibit improved trainability and fewer errors. The frequency of successful training in the pruned networks is comparable to smaller fully connected networks despite learning more mappings.

As the network size increases, training and controlling the network is more challenging. Measuring the trained state involves evaluating the effective {conductivity} between inputs and output nodes. In the effective circuit, current along an effective edge allows us to calculate the measurable current, 
\begin{align}
    \vec{i}_a = V_0 \tilde{G}_{0a} ,
\end{align} 
where output current through node $a$ is measured when voltage is applied to input node $0$, $\tilde{G}_{0a}$ is the effective conductivity between these nodes. The two-point effective conductance is {given by}
\begin{align}
    \tilde{G}_{0,a} &=  \sum_j \frac{G_{0,j}G_{a,j}}{\sum G_j} ,
\end{align}
where $\sum G_j$ is a sum over all edges connected to node $j$ in the hidden layer.

Consider the case of a network with two inputs $(0,1)$ and two outputs, $(a,b)$. To satisfy two mappings, $\left[0, a\right]$ and $ \left[1, b\right]$, the effective conductivity needs to satisfy 
\begin{subequations}
\begin{equation}
    \tilde{G}_{0a} -\tilde{G}_{0b} > 0 ,
 \end{equation}
\begin{equation}
   \sum_j \frac{G_{0j}(G_{ja}-G_{jb})}{\sum G_j} >0  ,
\end{equation}
    \label{eqn:EffectiveInequality1}
\end{subequations}
and 
\begin{subequations}
\begin{equation}
    \tilde{G}_{1a}  -\tilde{G}_{1b}<0 ,
 \end{equation}
\begin{equation}
    \sum_j \frac{G_{1j}(G_{ja}-G_{jb})}{\sum G_j} <0 .
    \end{equation}
    \label{eqn:EffectiveInequality2}
\end{subequations}

%

Here, the applied bias is omitted as it is constant for all mappings. Any two output currents in larger networks can be computed in this way. 
As all $G_{ij}$ are positive, to satisfy both inequalities, there must be a highly conductive input to a middle layer node, e.g., $G_{1j}$, that scales the outputs from the hidden layer with the desired contrast in resistance, e.g., $G_{ja} > G_{jb}$. {Given the observations of this section, it is important to mention why certain networks are less trainable than others. For this reason, we now discuss the notion of controllability.}




\subsection*{Controllability}

{The fact that} the number of errors would increase with the network size is not {immediately} obvious. During the training, each correction is applied to a slightly different circuit, as the connections from outputs to ground are adjusted for each specific correction. {To understand carefully the issue of controllability, we consider the projector formalism.} Additionally, when the learning parameter $\beta$ varies, the update $\dot{\vec{x}}$ is not an eigenfunction of the loop projection operator $\Omega_A$. The eigenfunctions of $\Omega_A$ are analyzed in the {Supplementary Material.} Consequently, while one might expect to control arbitrary parallel paths, a correction signal can still significantly impact parallel paths. {For context, the issue of state reachability for memristive networks was considered also in the context of reservoir computing \cite{Sheldon2022}, where it was found that because of Kirchhoff's laws not all states can be reached.}

\begin{figure}[h!] \centering
\includegraphics[width=1\textwidth]{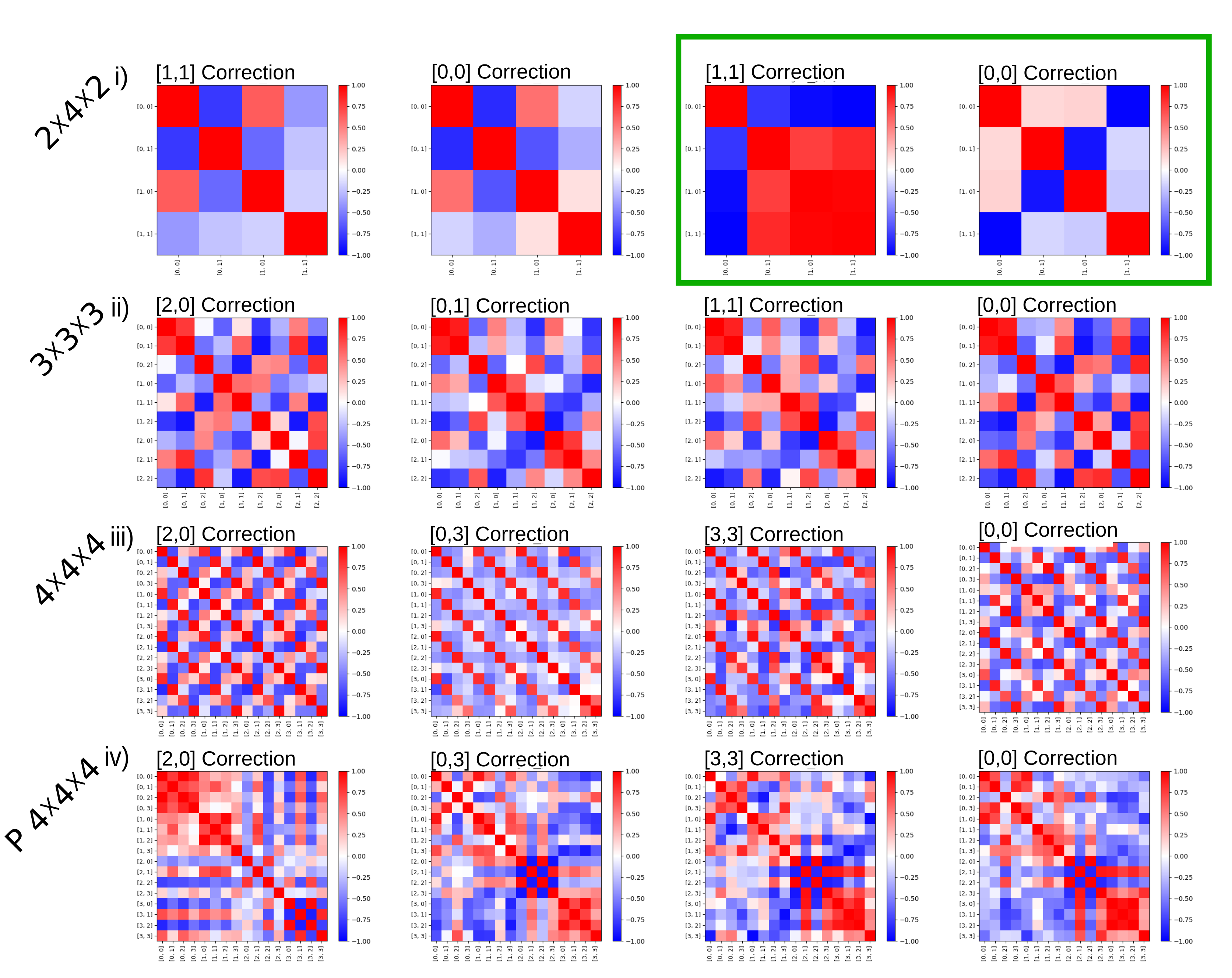}
\caption{Representative examples of correlations between two-point effective update functions throughout training in response to corrections. The two-point correlations are listed along the $x$ and $y$ axes of each plot, with each element representing the correlation between input and output nodes. For example, the top left element is $\text{Cor}(\dot{x}_{\left[0,0\right]},\dot{x}_{\left[0,0\right]})$ and the element below it is $\text{Cor}(\dot{x}_{\left[0,0\right]},\dot{x}_{\left[0,1\right]})$. Row (i): $2\times 4\times 2$ networks, row (ii): $3\times 3\times 3$ networks, (iii): the $4\times 4\times 4$ network, and row (iv): the pruned $4\times 4 \times 4$ network. Networks that successfully learn are highlighted with a green outline.  }
\label{fig:Correlations}
\end{figure}

Here we analyze correlations in the change {of the} effective conductivity. 
{Specifically, we} measure the two-point effective resistance between all combinations of input and output nodes following each to correction throughout training. 
Correlations {between the resistances} limit the ability to train the desired resistance contrast.
Representative examples of the correlations in an update of the two-point effective resistances, $\dot{\tilde{R}}$, are shown in Figure \ref{fig:Correlations}.
 These correlations illustrate how the effective resistance changes in response to corrections. In a fully {controllable} network, we would be able to independently update each effective resistance between any given input and output. In more complex networks, achieving this level of control is challenging.

In Figure \ref{fig:Correlations}, positive and negative correlations in effective resistance changes are shown in red and blue, respectively. These correlations persist over hundreds of corrections during training, with extensive positive and negative correlations observed between effective resistances in the network.
In the first row for the 2-input and 2-output networks, the first two columns show correlations in networks that do not successfully learn a pattern, while the third and fourth columns show correlations in networks that do learn a pattern.
Comparing the correlations in the 2-input 2-output networks in the top row, the off-diagonal positive correlations between the mappings $\left[0,1\right]$ and   $\left[1,0\right]$ in both correction signals hinders learning the correct mapping between $\left[1,0\right]$. In contrast, networks that successfully learn a pattern exhibit distinct correlations in response to distinct correction signals, with few off-diagonal elements showing similar positive or negative correlations in both signals.

In larger networks, correlations emerge between effective resistances sharing a common output node, leading to off-diagonal red diagonals. For the $3 \times 3 \times 3$ network, positive correlations are observed between effective resistances with common output nodes, along with block diagonal positive correlations from common input nodes. In the $4 \times 4 \times 4$ network, off-diagonal positive correlations are observed for resistances sharing a common output node, appearing in integer multiples of four above and below the main diagonal.
In comparison, the pruned 4-input and 4-output network shows fewer persistent correlations across all the corrections. Although a block diagonal structure is present in response to correction $\left[2,0\right]$, it is absent in other corrections. Similarly, the correlations between all input nodes with a common output node in the fully connected $4 \times 4 \times 4$ network are not observed in the pruned network.


\section*{Discussion}

Neuromorphic training on all-neuromorphic hardware can be effectively implemented using a learning-from-mistakes algorithm, demonstrating the potential of neuromorphic hardware combined with neuromorphic algorithms for efficient training. Our results show that even simple training rules are capable of training complex networks. The learning-from-mistakes algorithm facilitates self-organized learning by allowing for pruning throughout the network while only accessing the input and output layers. This practical approach avoids the need for intricate control over individual circuit elements and requires only access to the input and {output} layers. {This type of learning protocol is inspired by realistic biological protocols, in which actions occur only at the inputs and outputs nodes.}

Network co-design can improve the performance of neuromorphic training and decrease training errors. Co-design is possible with analytical tools to evaluate training performance. By calculating network capacity, effective conductivity, and correlations during training, we designed circuits with improved performance. While it is possible to design more controllable circuits, such as those with a fully connected input-output layer or CMOS-controlled {memristive devices}, our work focuses on quantifying and designing more complex networks with inaccessible hidden layers.

Two competing factors are involved in circuit design: controllability and capacity. Increasing network size improves capacity, as seen in {Figure \ref{fig:Capacities}}
 (a), but decreases controllability and trainability, resulting in higher training errors and correlations. 
{On the other hand, we observed that the} error rate increases with increasing network size and the number of hidden layer nodes. This is evidenced by the training success rates in Figure \ref{fig:AllError}, the sublinear growth in Figure \ref{fig:Capacities} (a), and the correlations in a larger network of Figure \ref{fig:Correlations}.  For example, $4\times6\times 4$ networks exhibit higher ensemble capacity compared to other fully connected $4$-in $4$-out networks but also have a higher error rate.
In addition, the success of training varies and depends on the specific network parameters, such as the range of $\beta$ values, as well as the random sequence of mappings that are trained. As such, the learning-from-mistakes algorithm is computationally non-deterministic. Despite this, {we have shown that} optimizing the hardware via co-design improves the performance of training {and the overall performance}.

Our key insight is in the role of network topology in balancing controllability and capacity. Specifically, cycles within the circuit impact correlations and interference, reducing the efficacy of correction signals. Networks with more cycles face challenges in learning multiple patterns due to overlapping conductive pathways, parallel paths, and leakage currents.
Let us now briefly discuss the differences between the present manuscript paper and the existing literature on the subject. First, we note that the BC algorithm introduced in \cite{Carbajal_2022} and based on \cite{Chialvo_1999}, and recently implemented in hardware \cite{Nikiruy_2024} {reinforces only the memristive devices on the shortest path, e.g. disconnecting parallel paths and using a modified punishment scheme that can be implemented on crossbars. 
We refer to Supplementary Material \ref{app:algorithmsimp} for an in-depth explanation. In this case, it is known that a larger middle layer implies improved learning, in contrast with what has been observed here. The present algorithm is, however, fully biological and physically plausible, similar to the algorithm for instance implemented in \cite{Loeffler2023} where there were also difficulties in the training scheme. This will be the focus of future studies.
Our findings are also in contrast with conventional (neural) network design as increasing the size of the hidden layer does improve network performance.

By calculating ensemble capacity and effective resistance, we can design network topologies that maintain connectivity and hidden layers while being easier to train.

Pruned networks are better at satisfying effective conductivity inequalities due to local resistance contrast. 
Simulations show that local resistance contrast, introduced through a general pruning mechanism, results in dominant pathways with high conductivity between input and output edges. Figure  \ref{fig:Correlations} highlights that pruned networks exhibit fewer correlations in response to corrections, allowing for better isolation and adjustment of specific two-point effective resistances throughout training.
The presence of more cycles in the circuit leads to correlations and interference {that} reduces the efficacy of correction signals.

Building upon this work larger all-{memristive} networks could be constructed from small controllable {memristive devices}. These functional units can be connected together into larger networks via nonlinear or thresholding elements. This is necessary as {memristive devices} do not function as perceptrons in circuits, to adapt an algorithm originally designed for perceptrons to {resistors with memory} one needs to adjust the algorithm or adjust the network; in contrast to previous works, we opted to adjust the network structure. It is important to stress that there exist neuromorphic-compatible models to build boolean functions (e.g. the receptron, see for instance \cite{Martini2022,Martini2024}).

Training learning-from-mistakes as implemented in the present manuscript is not optimally energy efficient.  Corrections required roughly 0.01 W though power consumption decreases throughout training, regardless of whether a pattern set is successfully learned. Once stabilized in a learned state the power consumption drops dramatically as it takes minimal amounts of energy to probe the memory of a nonvolatile {memristive} circuit.

Further work should focus on scaling these networks to larger configurations. Larger networks using small, controllable functional units could demonstrate the potential of an all-neuromorphic deep neural network.
The current study focused on small-scale networks as these are amenable to experimentation and analysis, thus scalability in generic networks remains an open question. Co-design will be crucial for linking small functional units to achieve desired functionality in larger networks. Additionally, innovative circuit topologies could further enhance controllability and capacity without relying on small functional units, as tuning topology remains an effective strategy for designing neuromorphic circuits.


\section*{Methods}
\subsection*{Experiments}

We experimentally implemented {memristive} networks using KnowM tungsten-doped DIPs, which contain eight {memristive devices} per chip {(pin to pin)}\cite{Campbell2017,Ostrovskii2021}. The network was biased using an SMU2600B voltage source, and current readings were taken with DMM6500 ammeters at each output. Hysteretic behavior was verified in selected pins using SMU2600B IV characterization, where each pin was biased with two periods of a triangle wave sweep with an amplitude of \SI{2}{\volt} and initial phase \SI{0}{\volt}. To avoid {damaging} the {devices}, all voltages were applied with a maximum current of \SI{60}{\micro\ampere}. Pins exhibiting clear hysteretic behavior were selected for inclusion in the network. Ultimately, 16 pins were selected for the final $2 \times 4 \times 2$ topology.

Networks were trained to switch between two pattern sets: the ``identity" set $\{(1, 0) \mapsto (1, 0), (0, 1) \mapsto (0, 1)\}$ and the "rotation" set $\{(0, 1) \mapsto (1, 0), (1, 0) \mapsto (0, 1)\}$. A single ``epoch" of training comprised 40 random samplings from a given pattern set, after which the active pattern set was switched. The network thus had 40 chances to learn a given pattern set before swaps. Each random sampling corresponded to a training step, in which the current network map was read and corrected if necessary. Performance was evaluated over 10 swaps or 20 total pattern sets, a total of 800 samplings. 

Read biases were applied as low-voltage \SI{20}{\micro\second} square waves to avoid overly perturbing the network state. Output currents had magnitudes ranging from \SI{0.1}{\micro\ampere} to \SI{10}{\nano\ampere}. Generally, output currents had magnitude differences ranging from \SI{12}{\nano\ampere} to \SI{200}{\nano\ampere}. Smaller differences were considered spurious. In these cases, the magnitude of read voltage was increased geometrically until differences increased to an acceptable magnitude, beginning at \SI{0.05}{\volt} and increasing by a factor of 1.2 to a maximum \SI{0.1}{\volt}. In the very rare case that differences remained small at this point, a random output pathway was chosen for correction.

During training, corrective signals comprised negative correction and positive normalization biasing. A gating mechanism is required during the negative correction steps, in which the correct/target output node is disconnected totally from {the} ground. In practice, it was found that simply switching the DMM6500 ammeter at this output to a high-impedance \SI{10}{\mega\ohm} state sufficiently differentiated the two output pathways to allow learning. Correction biases were applied as \SI{3}{\second} linear ramps from 0 to \SI{-4}{\volt}. Normalization biases, which were performed immediately after each correction bias to avoid saturating {the devices'} memory values, were \SI{1}{\second} linear ramps from 0 to \SI{0.2}{\volt} with both output nodes connected to ground. The current cap was again set to \SI{60}{\micro\ampere} during these biasing.

\subsection*{Simulations}
Simulations were conducted using the PySpice package\cite{PySpice} and numerical integration, yielding results consistent with experimental observations. The PySpice simulations utilized a nonlinear {memristive device} model.\cite{Yakopcic_2012} {The linear memristive devices} were modeled through numerical integration based on equation \ref{eqn:CaravelliEqn}.\cite{caravelli2017complex} 
In SPICE simulations, gates were used to adjust the circuit such that corrective signals biased the desired output node and only the erroneous output node was connected to {the ground}. In numerical integration, corrective signals were applied to circuits with distinct {connectivity}. Simulations using equation \ref{eqn:CaravelliEqn} were comparable to experiments and PySpice simulations, and {scaled} to larger networks better than PySpice.

{The training} involved two alternating pattern sets across three training eras.  Each training era consisted of {a} $500$ random sampling of the mappings from a pattern. If the network had not learned a mapping after 500 steps, it was considered unsuccessful in learning that mapping. Epochs for the simulated networks {are} defined as $50$ random mappings, with each mapping read into the network using a low magnitude read bias. If the read operation produced an incorrect output, a correction and normalization bias were applied as described in the main text.

The resistance states in the simulations varied by three orders of magnitude: the ``on" state had resistance values between $10$ Ohms and $1$ kOhms, while the ``off" state ranged from $10$ kOhms to $1$ MOhms. Measured output currents were considered distinct if they differed by more than $20$ nA.

{The bias} applied during read operations ranged between $0.1 - 1$ mV.  Correction and normalization biases varied by network size. For small $2$-input $2$-output networks the magnitude of the correction bias was $0.25$ Volts,  increasing proportionally with the number of input nodes in larger networks. The normalization bias was typically {one-half} of the correction bias. In the pruned networks the normalization signal was one-quarter the correction signal. Consistent read, correction, and normalization biases were used for networks of the same size. 

For both the linear and nonlinear {memristive devices}, the learning rate for each {memristive device} was randomly sampled from a range of $0.05$ - $0.15$ $V\cdot s$ at the start of training.  Initial memory parameters $x_0$ were randomly sampled from a range of $0.2$ - $0.8$. The volatility, $\alpha$, for {linear devices} is assumed to be negligible.

Average training results, as shown in Figures \ref{fig:IntegrationCNN} and \ref{fig:AllError} were obtained from multiple training sessions with different initial networks. Each network size was tested with 20 different initial configurations. The average capacity depicted in Figure \ref{fig:Capacities} was determined for networks with random resistance values, using equations \ref{eqn:StatMechCapacity}. Ensemble capacity was determined over $1000$ network configurations.

\section *{Acknowledgements}
The authors acknowledge the support of the NNSA for
the U.S. DoE at LANL under Contract No. DE-AC52-
06NA25396, and Laboratory Directed Research and Development (LDRD). DRC acknowledges support from the NIH Brain Initiative (USA) Grant No.
1U19NS107464-01. FC and JL were financed via DOE LDRD grant 20240245ER, while FB via Director’s Fellowship. FB and JL also gratefully acknowledge support from the Center for Nonlinear Studies at LANL and the Center for Integrated Nanotechnology where these experiments were conducted. We are in particular indebted to Aiping Chen, and Michael Shlanta for help in setting up the experiments at CINT, and in particular Magdalena E. Dale and Michael Isaf for an earlier characterization of the devices we used in unpublished work.

\section*{Author contributions}
F. Barrows performed the numerical work. J. Lin and F. Barrows performed experimental work in this order. F. Barrows and F. Caravelli performed analytical work in this order. D.R. Chialvo and F. Caravelli conceived the initial idea for the project based on earlier work from the former. F. Barrows wrote the initial draft. 
 All authors contributed to the writing of the final draft.

\bibliographystyle{naturemag}
\bibliography{bibliography}

\begin{thebibliography}{10}
\expandafter\ifx\csname url\endcsname\relax
  \def\url#1{\texttt{#1}}\fi
\expandafter\ifx\csname urlprefix\endcsname\relax\def\urlprefix{URL }\fi
\providecommand{\bibinfo}[2]{#2}
\providecommand{\eprint}[2][]{\url{#2}}

\bibitem{Engelhardt2021-ic}
\bibinfo{author}{Engelhardt, E.}
\newblock \bibinfo{title}{Descartes and his project of a fantasized brain}.
\newblock \emph{\bibinfo{journal}{Dementia \& Neuropsychologia}} \textbf{\bibinfo{volume}{15}}, \bibinfo{pages}{281–285} (\bibinfo{year}{2021}).
\newblock \urlprefix\url{http://dx.doi.org/10.1590/1980-57642021dn15-020017}.

\bibitem{Turing_1950}
\bibinfo{author}{TURING, A.~M.}
\newblock \bibinfo{title}{I.—computing machinery and intelligence}.
\newblock \emph{\bibinfo{journal}{Mind}} \textbf{\bibinfo{volume}{LIX}}, \bibinfo{pages}{433–460} (\bibinfo{year}{1950}).
\newblock \urlprefix\url{http://dx.doi.org/10.1093/mind/LIX.236.433}.

\bibitem{Turing_1952}
\bibinfo{author}{Turing, A.} \& \bibinfo{author}{Braithwaite, R.}
\newblock \emph{\bibinfo{title}{Can Automatic Calculating Machines Be Said To Think? (1952)}}, \bibinfo{pages}{487–506} (\bibinfo{publisher}{Oxford University PressOxford}, \bibinfo{year}{2004}).
\newblock \urlprefix\url{http://dx.doi.org/10.1093/oso/9780198250791.003.0020}.

\bibitem{Miller_2003}
\bibinfo{author}{Miller, G.~A.}
\newblock \bibinfo{title}{The cognitive revolution: a historical perspective}.
\newblock \emph{\bibinfo{journal}{Trends in Cognitive Sciences}} \textbf{\bibinfo{volume}{7}}, \bibinfo{pages}{141--144} (\bibinfo{year}{2003}).
\newblock \urlprefix\url{https://doi.org/10.1016/S1364-6613(03)00029-9}.

\bibitem{Fjelland_2020}
\bibinfo{author}{Fjelland, R.}
\newblock \bibinfo{title}{Why general artificial intelligence will not be realized}.
\newblock \emph{\bibinfo{journal}{Humanities and Social Sciences Communications}} \textbf{\bibinfo{volume}{7}}, \bibinfo{pages}{10} (\bibinfo{year}{2020}).
\newblock \urlprefix\url{https://doi.org/10.1057/s41599-020-0494-4}.

\bibitem{George_2022}
\bibinfo{author}{Deane, G.}
\newblock \bibinfo{title}{{Machines That Feel and Think: The Role of Affective Feelings and Mental Action in (Artificial) General Intelligence}}.
\newblock \emph{\bibinfo{journal}{Artificial Life}} \textbf{\bibinfo{volume}{28}}, \bibinfo{pages}{289--309} (\bibinfo{year}{2022}).
\newblock \urlprefix\url{https://doi.org/10.1162/artl\_a\_00368}.
\newblock \eprint{https://direct.mit.edu/artl/article-pdf/28/3/289/2037982/artl\_a\_00368.pdf}.

\bibitem{bubeck2023sparksartificialgeneralintelligence}
\bibinfo{author}{Bubeck, S.} \emph{et~al.}
\newblock \bibinfo{title}{Sparks of artificial general intelligence: Early experiments with gpt-4} (\bibinfo{year}{2023}).
\newblock \urlprefix\url{https://arxiv.org/abs/2303.12712}.
\newblock \eprint{2303.12712}.

\bibitem{Menabrea_2015}
\bibinfo{author}{Menabrea, L.~F.}
\newblock \emph{\bibinfo{title}{Sketch of the Analytical Engine invented by Charles Babbage, Esq.}} (\bibinfo{publisher}{Association for Computing Machinery and Morgan \& Claypool}, \bibinfo{year}{2015}).
\newblock \urlprefix\url{https://doi.org/10.1145/2809523.2809528}.

\bibitem{Schuman_2022}
\bibinfo{author}{Schuman, C.~D.} \emph{et~al.}
\newblock \bibinfo{title}{Opportunities for neuromorphic computing algorithms and applications}.
\newblock \emph{\bibinfo{journal}{Nature Computational Science}} \textbf{\bibinfo{volume}{2}}, \bibinfo{pages}{10--19} (\bibinfo{year}{2022}).
\newblock \urlprefix\url{https://doi.org/10.1038/s43588-021-00184-y}.

\bibitem{Christensen_2022}
\bibinfo{author}{Christensen, D.~V.} \emph{et~al.}
\newblock \bibinfo{title}{2022 roadmap on neuromorphic computing and engineering}.
\newblock \emph{\bibinfo{journal}{Neuromorphic Computing and Engineering}} \textbf{\bibinfo{volume}{2}}, \bibinfo{pages}{022501} (\bibinfo{year}{2022}).
\newblock \urlprefix\url{https://dx.doi.org/10.1088/2634-4386/ac4a83}.

\bibitem{Kim_Front_2024}
\bibinfo{author}{Kim, K.}, \bibinfo{author}{Song, M.~S.}, \bibinfo{author}{Hwang, H.}, \bibinfo{author}{Hwang, S.} \& \bibinfo{author}{Kim, H.}
\newblock \bibinfo{title}{A comprehensive review of advanced trends: from artificial synapses to neuromorphic systems with consideration of non-ideal effects}.
\newblock \emph{\bibinfo{journal}{Frontiers in Neuroscience}} \textbf{\bibinfo{volume}{18}} (\bibinfo{year}{2024}).
\newblock \urlprefix\url{https://www.frontiersin.org/journals/neuroscience/articles/10.3389/fnins.2024.1279708}.

\bibitem{Chialvo_1999}
\bibinfo{author}{Chialvo, D.} \& \bibinfo{author}{Bak, P.}
\newblock \bibinfo{title}{Learning from mistakes}.
\newblock \emph{\bibinfo{journal}{Neuroscience}} \textbf{\bibinfo{volume}{90}}, \bibinfo{pages}{1137--1148} (\bibinfo{year}{1999}).
\newblock \urlprefix\url{https://www.sciencedirect.com/science/article/pii/S0306452298004722}.

\bibitem{Bak_2001}
\bibinfo{author}{Bak, P.} \& \bibinfo{author}{Chialvo, D.~R.}
\newblock \bibinfo{title}{Adaptive learning by extremal dynamics and negative feedback}.
\newblock \emph{\bibinfo{journal}{Phys. Rev. E}} \textbf{\bibinfo{volume}{63}}, \bibinfo{pages}{031912} (\bibinfo{year}{2001}).
\newblock \urlprefix\url{https://link.aps.org/doi/10.1103/PhysRevE.63.031912}.

\bibitem{Wakeling_2003}
\bibinfo{author}{Wakeling, J.}
\newblock \bibinfo{title}{Order–disorder transition in the chialvo–bak ‘minibrain’ controlled by network geometry}.
\newblock \emph{\bibinfo{journal}{Physica A: Statistical Mechanics and its Applications}} \textbf{\bibinfo{volume}{325}}, \bibinfo{pages}{561--569} (\bibinfo{year}{2003}).
\newblock \urlprefix\url{https://www.sciencedirect.com/science/article/pii/S037843710300147X}.

\bibitem{Brigham_2009}
\bibinfo{author}{Brigham, M.}
\newblock \emph{\bibinfo{title}{Self-Organised Learning in the Chialvo-Bak Model}}.
\newblock Ph.D. thesis, \bibinfo{school}{University of Edinburgh} (\bibinfo{year}{2009}).

\bibitem{Carbajal_2022}
\bibinfo{author}{Carbajal, J.~P.}, \bibinfo{author}{Martin, D.~A.} \& \bibinfo{author}{Chialvo, D.~R.}
\newblock \bibinfo{title}{Learning by mistakes in memristor networks}.
\newblock \emph{\bibinfo{journal}{Phys. Rev. E}} \textbf{\bibinfo{volume}{105}}, \bibinfo{pages}{054306} (\bibinfo{year}{2022}).
\newblock \urlprefix\url{https://link.aps.org/doi/10.1103/PhysRevE.105.054306}.

\bibitem{Nikiruy_2024}
\bibinfo{author}{Nikiruy, K.} \emph{et~al.}
\newblock \bibinfo{title}{Blooming and pruning: learning from mistakes with memristive synapses}.
\newblock \emph{\bibinfo{journal}{Scientific Reports}} \textbf{\bibinfo{volume}{14}}, \bibinfo{pages}{7802} (\bibinfo{year}{2024}).
\newblock \urlprefix\url{https://doi.org/10.1038/s41598-024-57660-4}.

\bibitem{Hooker_2021}
\bibinfo{author}{Hooker, S.}
\newblock \bibinfo{title}{The hardware lottery}.
\newblock \emph{\bibinfo{journal}{Commun. ACM}} \textbf{\bibinfo{volume}{64}}, \bibinfo{pages}{58–65} (\bibinfo{year}{2021}).
\newblock \urlprefix\url{https://doi.org/10.1145/3467017}.

\bibitem{MONTAGU_1964}
\bibinfo{author}{MONTAGU, A.}
\newblock \bibinfo{title}{The postnatal development of the human cerebral cortex}.
\newblock \emph{\bibinfo{journal}{American Journal of Psychiatry}} \textbf{\bibinfo{volume}{120}}, \bibinfo{pages}{933--933} (\bibinfo{year}{1964}).
\newblock \urlprefix\url{https://doi.org/10.1176/ajp.120.9.933}.
\newblock \eprint{https://doi.org/10.1176/ajp.120.9.933}.

\bibitem{CHECHIK_1999}
\bibinfo{author}{Chechik, G.}, \bibinfo{author}{Meilijson, I.} \& \bibinfo{author}{Ruppin, E.}
\newblock \bibinfo{title}{Neuronal regulation: A biologically plausible mechanism for efficient synaptic pruning in development}.
\newblock \emph{\bibinfo{journal}{Neurocomputing}} \textbf{\bibinfo{volume}{26-27}}, \bibinfo{pages}{633--639} (\bibinfo{year}{1999}).
\newblock \urlprefix\url{https://www.sciencedirect.com/science/article/pii/S0925231298001611}.

\bibitem{Faust_2021}
\bibinfo{author}{Faust, T.~E.}, \bibinfo{author}{Gunner, G.} \& \bibinfo{author}{Schafer, D.~P.}
\newblock \bibinfo{title}{Mechanisms governing activity-dependent synaptic pruning in the developing mammalian cns}.
\newblock \emph{\bibinfo{journal}{Nature Reviews Neuroscience}} \textbf{\bibinfo{volume}{22}}, \bibinfo{pages}{657--673} (\bibinfo{year}{2021}).
\newblock \urlprefix\url{https://doi.org/10.1038/s41583-021-00507-y}.

\bibitem{Sanger_1989}
\bibinfo{author}{Sanger, T.~D.}
\newblock \bibinfo{title}{Optimal unsupervised learning in a single-layer linear feedforward neural network}.
\newblock \emph{\bibinfo{journal}{Neural Networks}} \textbf{\bibinfo{volume}{2}}, \bibinfo{pages}{459--473} (\bibinfo{year}{1989}).
\newblock \urlprefix\url{https://www.sciencedirect.com/science/article/pii/0893608089900440}.

\bibitem{Friston_2010}
\bibinfo{author}{Friston, K.}
\newblock \bibinfo{title}{The free-energy principle: a unified brain theory?}
\newblock \emph{\bibinfo{journal}{Nature Reviews Neuroscience}} \textbf{\bibinfo{volume}{11}}, \bibinfo{pages}{127--138} (\bibinfo{year}{2010}).
\newblock \urlprefix\url{https://doi.org/10.1038/nrn2787}.

\bibitem{rosenblatt1961principles}
\bibinfo{author}{Rosenblatt, F.}
\newblock \bibinfo{title}{Principles of neurodynamics. perceptrons and the theory of brain mechanisms}.
\newblock \bibinfo{type}{Tech. Rep.}, \bibinfo{institution}{Cornell Aeronautical Lab Inc Buffalo NY} (\bibinfo{year}{1961}).

\bibitem{Linnainmaa_1976}
\bibinfo{author}{Linnainmaa, S.}
\newblock \bibinfo{title}{Taylor expansion of the accumulated rounding error}.
\newblock \emph{\bibinfo{journal}{BIT Numerical Mathematics}} \textbf{\bibinfo{volume}{16}}, \bibinfo{pages}{146--160} (\bibinfo{year}{1976}).
\newblock \urlprefix\url{https://doi.org/10.1007/BF01931367}.

\bibitem{Scellier_2017}
\bibinfo{author}{Scellier, B.} \& \bibinfo{author}{Bengio, Y.}
\newblock \bibinfo{title}{Equilibrium propagation: Bridging the gap between energy-based models and backpropagation}.
\newblock \emph{\bibinfo{journal}{Frontiers in Computational Neuroscience}} \textbf{\bibinfo{volume}{11}} (\bibinfo{year}{2017}).
\newblock \urlprefix\url{https://www.frontiersin.org/journals/computational-neuroscience/articles/10.3389/fncom.2017.00024}.

\bibitem{Block_1962}
\bibinfo{author}{Block, H.~D.}
\newblock \bibinfo{title}{The perceptron: A model for brain functioning. i}.
\newblock \emph{\bibinfo{journal}{Rev. Mod. Phys.}} \textbf{\bibinfo{volume}{34}}, \bibinfo{pages}{123--135} (\bibinfo{year}{1962}).
\newblock \urlprefix\url{https://link.aps.org/doi/10.1103/RevModPhys.34.123}.

\bibitem{rosenblatt1957perceptron}
\bibinfo{author}{Rosenblatt, F.}
\newblock \emph{\bibinfo{title}{The Perceptron, a Perceiving and Recognizing Automaton Project Para}}.
\newblock Report: Cornell Aeronautical Laboratory (\bibinfo{publisher}{Cornell Aeronautical Laboratory}, \bibinfo{year}{1957}).
\newblock \urlprefix\url{https://books.google.com/books?id=P_XGPgAACAAJ}.

\bibitem{Silva_2020}
\bibinfo{author}{Silva, F.}, \bibinfo{author}{Sanz, M.}, \bibinfo{author}{Seixas, J.}, \bibinfo{author}{Solano, E.} \& \bibinfo{author}{Omar, Y.}
\newblock \bibinfo{title}{Perceptrons from memristors}.
\newblock \emph{\bibinfo{journal}{Neural Networks}} \textbf{\bibinfo{volume}{122}}, \bibinfo{pages}{273--278} (\bibinfo{year}{2020}).
\newblock \urlprefix\url{https://www.sciencedirect.com/science/article/pii/S0893608019303351}.

\bibitem{Strukov_2008}
\bibinfo{author}{Strukov, D.~B.}, \bibinfo{author}{Snider, G.~S.}, \bibinfo{author}{Stewart, D.~R.} \& \bibinfo{author}{Williams, R.~S.}
\newblock \bibinfo{title}{The missing memristor found}.
\newblock \emph{\bibinfo{journal}{Nature}} \textbf{\bibinfo{volume}{453}}, \bibinfo{pages}{80--83} (\bibinfo{year}{2008}).
\newblock \urlprefix\url{https://doi.org/10.1038/nature06932}.

\bibitem{reviewCarCar}
\bibinfo{author}{Caravelli, F.} \& \bibinfo{author}{Carbajal, J.~P.}
\newblock \bibinfo{title}{Memristors for the curious outsiders}.
\newblock \emph{\bibinfo{journal}{Technologies}} \textbf{\bibinfo{volume}{6}}, \bibinfo{pages}{118} (\bibinfo{year}{2018}).

\bibitem{Kumar_2022}
\bibinfo{author}{Kumar, S.}, \bibinfo{author}{Wang, X.}, \bibinfo{author}{Strachan, J.~P.}, \bibinfo{author}{Yang, Y.} \& \bibinfo{author}{Lu, W.~D.}
\newblock \bibinfo{title}{Dynamical memristors for higher-complexity neuromorphic computing}.
\newblock \emph{\bibinfo{journal}{Nature Reviews Materials}} \textbf{\bibinfo{volume}{7}}, \bibinfo{pages}{575--591} (\bibinfo{year}{2022}).
\newblock \urlprefix\url{https://doi.org/10.1038/s41578-022-00434-z}.

\bibitem{Xiao_2023}
\bibinfo{author}{Xiao, Y.} \emph{et~al.}
\newblock \bibinfo{title}{A review of memristor: material and structure design, device performance, applications and prospects}.
\newblock \emph{\bibinfo{journal}{Science and Technology of Advanced Materials}} \textbf{\bibinfo{volume}{24}} (\bibinfo{year}{2023}).
\newblock \urlprefix\url{http://dx.doi.org/10.1080/14686996.2022.2162323}.

\bibitem{Mambretti2022}
\bibinfo{author}{Mambretti, F.} \emph{et~al.}
\newblock \bibinfo{title}{Dynamical stochastic simulation of complex electrical behavior in neuromorphic networks of metallic nanojunctions}.
\newblock \emph{\bibinfo{journal}{Scientific Reports}} \textbf{\bibinfo{volume}{12}} (\bibinfo{year}{2022}).
\newblock \urlprefix\url{http://dx.doi.org/10.1038/s41598-022-15996-9}.

\bibitem{Gaba_2013}
\bibinfo{author}{Gaba, S.}, \bibinfo{author}{Sheridan, P.}, \bibinfo{author}{Zhou, J.}, \bibinfo{author}{Choi, S.} \& \bibinfo{author}{Lu, W.}
\newblock \bibinfo{title}{Stochastic memristive devices for computing and neuromorphic applications}.
\newblock \emph{\bibinfo{journal}{Nanoscale}} \textbf{\bibinfo{volume}{5}}, \bibinfo{pages}{5872--5878} (\bibinfo{year}{2013}).
\newblock \urlprefix\url{http://dx.doi.org/10.1039/C3NR01176C}.

\bibitem{Ignatov_2017}
\bibinfo{author}{Ignatov, M.}, \bibinfo{author}{Ziegler, M.}, \bibinfo{author}{Hansen, M.} \& \bibinfo{author}{Kohlstedt, H.}
\newblock \bibinfo{title}{Memristive stochastic plasticity enables mimicking of neural synchrony: Memristive circuit emulates an optical illusion}.
\newblock \emph{\bibinfo{journal}{Science Advances}} \textbf{\bibinfo{volume}{3}}, \bibinfo{pages}{e1700849} (\bibinfo{year}{2017}).
\newblock \urlprefix\url{https://www.science.org/doi/abs/10.1126/sciadv.1700849}.
\newblock \eprint{https://www.science.org/doi/pdf/10.1126/sciadv.1700849}.

\bibitem{Lin_2024}
\bibinfo{author}{Lin, J.}, \bibinfo{author}{Barrows, F.} \& \bibinfo{author}{Caravelli, F.}
\newblock \bibinfo{title}{Memristive linear algebra}.
\newblock \emph{\bibinfo{journal}{arXiv preprint arXiv:2407.20539}}  (\bibinfo{year}{2024}).
\newblock \urlprefix\url{https://arxiv.org/abs/2407.20539}.

\bibitem{Mirigliano2021}
\bibinfo{author}{Mirigliano, M.} \emph{et~al.}
\newblock \bibinfo{title}{A binary classifier based on a reconfigurable dense network of metallic nanojunctions}.
\newblock \emph{\bibinfo{journal}{Neuromorphic Computing and Engineering}} \textbf{\bibinfo{volume}{1}}, \bibinfo{pages}{024007} (\bibinfo{year}{2021}).
\newblock \urlprefix\url{http://dx.doi.org/10.1088/2634-4386/ac29c9}.

\bibitem{PySpice}
\bibinfo{author}{Salvaire, F.}
\newblock \bibinfo{title}{Pyspice}.
\newblock \urlprefix\url{https://pyspice.fabrice-salvaire.fr}.

\bibitem{caravelli2017complex}
\bibinfo{author}{Caravelli, F.}, \bibinfo{author}{Traversa, F.~L.} \& \bibinfo{author}{Di~Ventra, M.}
\newblock \bibinfo{title}{Complex dynamics of memristive circuits: Analytical results and universal slow relaxation}.
\newblock \emph{\bibinfo{journal}{Physical Review E}} \textbf{\bibinfo{volume}{95}}, \bibinfo{pages}{022140} (\bibinfo{year}{2017}).

\bibitem{Caravellisciad}
\bibinfo{author}{Caravelli, F.}, \bibinfo{author}{Sheldon, F.~C.} \& \bibinfo{author}{Traversa, F.~L.}
\newblock \bibinfo{title}{Global minimization via classical tunneling assisted by collective force field formation}.
\newblock \emph{\bibinfo{journal}{Science Advances}} \textbf{\bibinfo{volume}{7}} (\bibinfo{year}{2021}).
\newblock \urlprefix\url{http://dx.doi.org/10.1126/sciadv.abh1542}.

\bibitem{barrows_2024}
\bibinfo{author}{Barrows, F.}, \bibinfo{author}{Sheldon, F.~C.} \& \bibinfo{author}{Caravelli, F.}
\newblock \bibinfo{title}{Network analysis of memristive device circuits: dynamics, stability and correlations}  (\bibinfo{year}{2024}).
\newblock \urlprefix\url{https://arxiv.org/abs/2402.16015}.
\newblock \eprint{2402.16015}.

\bibitem{Sheldon2022}
\bibinfo{author}{Sheldon, F.~C.}, \bibinfo{author}{Kolchinsky, A.} \& \bibinfo{author}{Caravelli, F.}
\newblock \bibinfo{title}{Computational capacity of lrc, memristive, and hybrid reservoirs}.
\newblock \emph{\bibinfo{journal}{Physical Review E}} \textbf{\bibinfo{volume}{106}} (\bibinfo{year}{2022}).
\newblock \urlprefix\url{http://dx.doi.org/10.1103/PhysRevE.106.045310}.

\bibitem{Loeffler2023}
\bibinfo{author}{Loeffler, A.} \emph{et~al.}
\newblock \bibinfo{title}{Neuromorphic learning, working memory, and metaplasticity in nanowire networks}.
\newblock \emph{\bibinfo{journal}{Science Advances}} \textbf{\bibinfo{volume}{9}} (\bibinfo{year}{2023}).
\newblock \urlprefix\url{http://dx.doi.org/10.1126/sciadv.adg3289}.

\bibitem{Martini2022}
\bibinfo{author}{Martini, G.}, \bibinfo{author}{Mirigliano, M.}, \bibinfo{author}{Paroli, B.} \& \bibinfo{author}{Milani, P.}
\newblock \bibinfo{title}{The receptron: a device for the implementation of information processing systems based on complex nanostructured systems}.
\newblock \emph{\bibinfo{journal}{Japanese Journal of Applied Physics}} \textbf{\bibinfo{volume}{61}}, \bibinfo{pages}{SM0801} (\bibinfo{year}{2022}).
\newblock \urlprefix\url{http://dx.doi.org/10.35848/1347-4065/ac665c}.

\bibitem{Martini2024}
\bibinfo{author}{Martini, G.} \emph{et~al.}
\newblock \bibinfo{title}{Efficiency and controllability of stochastic boolean function generation by a random network of non-linear nanoparticle junctions}.
\newblock \emph{\bibinfo{journal}{Frontiers in Physics}} \textbf{\bibinfo{volume}{12}} (\bibinfo{year}{2024}).
\newblock \urlprefix\url{http://dx.doi.org/10.3389/fphy.2024.1400919}.

\bibitem{Campbell2017}
\bibinfo{author}{Campbell, K.~A.}
\newblock \bibinfo{title}{Self-directed channel memristor for high temperature operation}.
\newblock \emph{\bibinfo{journal}{Microelectronics Journal}} \textbf{\bibinfo{volume}{59}}, \bibinfo{pages}{10–14} (\bibinfo{year}{2017}).
\newblock \urlprefix\url{http://dx.doi.org/10.1016/j.mejo.2016.11.006}.

\bibitem{Ostrovskii2021}
\bibinfo{author}{Ostrovskii, V.}, \bibinfo{author}{Fedoseev, P.}, \bibinfo{author}{Bobrova, Y.} \& \bibinfo{author}{Butusov, D.}
\newblock \bibinfo{title}{Structural and parametric identification of knowm memristors}.
\newblock \emph{\bibinfo{journal}{Nanomaterials}} \textbf{\bibinfo{volume}{12}}, \bibinfo{pages}{63} (\bibinfo{year}{2021}).
\newblock \urlprefix\url{http://dx.doi.org/10.3390/nano12010063}.

\bibitem{Yakopcic_2012}
\bibinfo{author}{Yakopcic, C.}, \bibinfo{author}{Taha, T.~M.}, \bibinfo{author}{Subramanyam, G.} \& \bibinfo{author}{Pino, R.~E.}
\newblock \emph{\bibinfo{title}{Memristor SPICE Modeling}}, \bibinfo{pages}{211--244} (\bibinfo{publisher}{Springer Netherlands}, \bibinfo{address}{Dordrecht}, \bibinfo{year}{2012}).
\newblock \urlprefix\url{https://doi.org/10.1007/978-94-007-4491-2_12}.

\bibitem{chialvo_bak_learning_1}
\bibinfo{author}{Carbajal, J.~P.}
\newblock \bibinfo{title}{Learning from mistakes 1} (\bibinfo{year}{2024}).
\newblock \urlprefix\url{https://kakila.gitlab.io/mistake-learning/mem-chialvo-bak.html}.
\newblock \bibinfo{note}{Accessed: 2024-08-01}.

\bibitem{chialvo_bak_learning_2}
\bibinfo{author}{Carbajal, J.~P.}
\newblock \bibinfo{title}{Learning from mistakes 2} (\bibinfo{year}{2024}).
\newblock \urlprefix\url{https://kakila.gitlab.io/mistake-learning/}.
\newblock \bibinfo{note}{Accessed: 2024-08-01}.

\end{thebibliography}

\clearpage
\appendix
\section{Supplementary Material}
\subsection{Training experimental network}
The hardware {memristive} neural network is shown in Figure \ref{fig:Breadboard}. The network consists of KNOWM {memristive devices} connected by wires. The junctions where the wires intersect serve as the nodes of the network. Voltage generators and ammeters are not depicted.
\begin{figure}[h!] \centering
    \includegraphics[width=.45\textwidth]{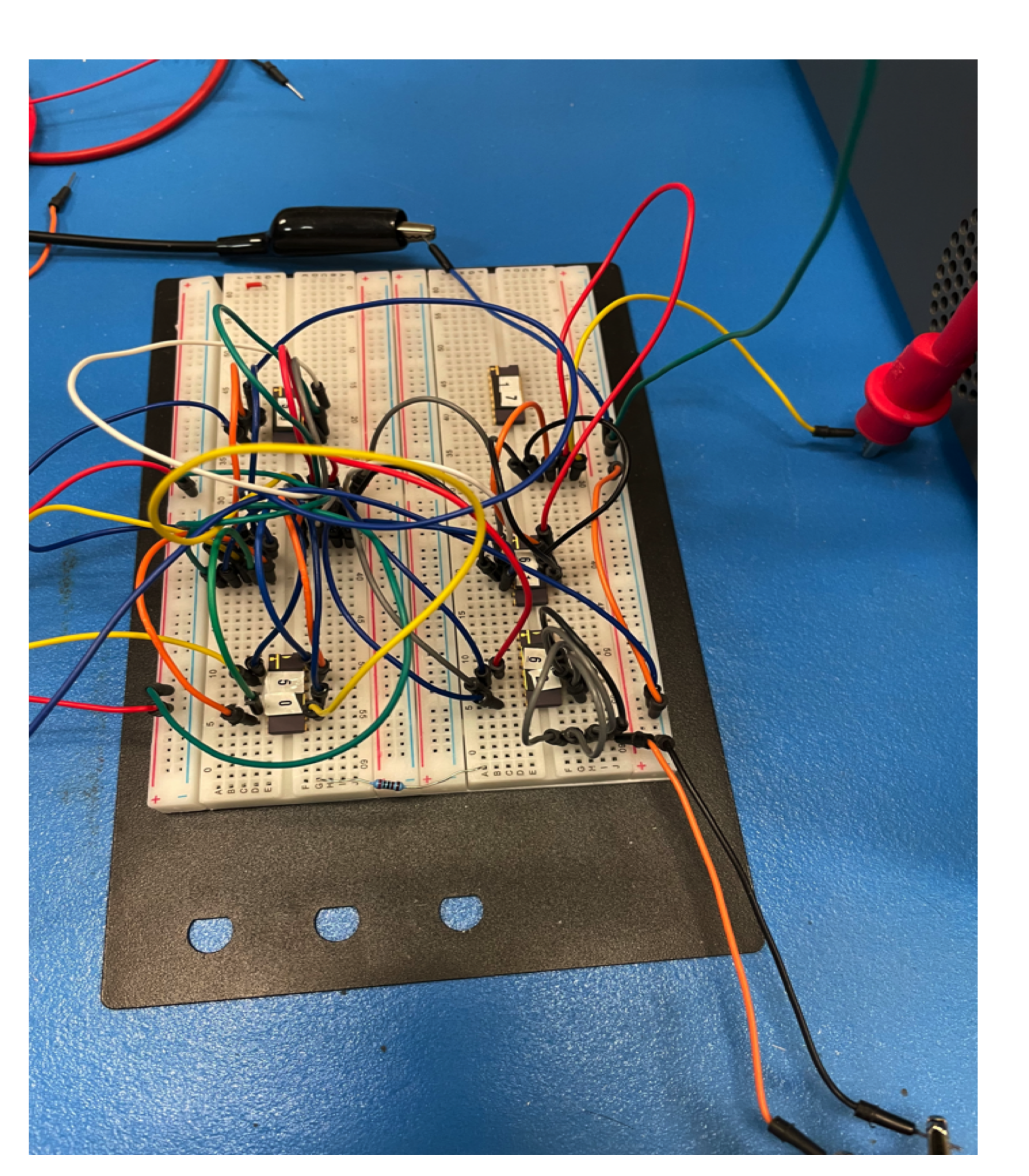}
    \caption{Photo of the {memristive} network. {The devices} are connected by wires on a breadboard, {and the memristive devices are pin-to-pin on each 16-pin chip}. Voltage generators and ammeters are not pictured. }
         \label{fig:Breadboard}
\end{figure}

 Figure \ref{fig:ExpAvg} presents the average error during the training of the experimental network. The experimental network consists of $2$-input nodes, $2$-output nodes, and $4$-nodes in the hidden layer, with full connectivity between subsequent layers.
\begin{figure}[h!] \centering
    \includegraphics[width=1\textwidth]{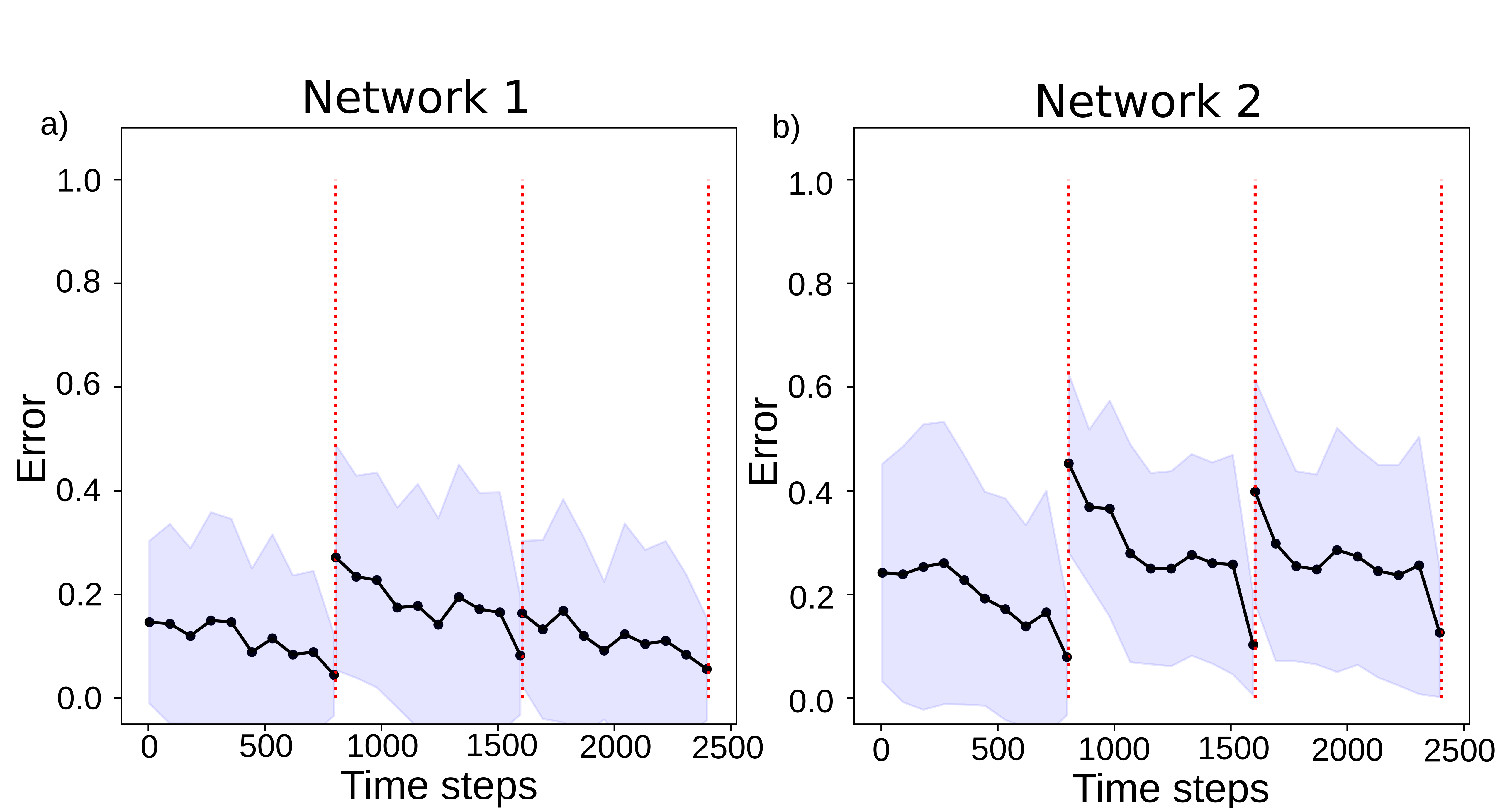}
    \caption{Average error during training an experimental $2\times4\times2$ network shown as a black line, with the standard deviation indicated by the shaded region. Two distinct patterns are trained over three training intervals, shown in (a) and (b) respectively. }
         \label{fig:ExpAvg}
\end{figure}
Two different $2\times4\times2$ networks were trained on two distinct patterns over three training eras, e.g. pattern 1, pattern 2, and pattern 1 were trained in sequence. This process was repeated eight times, with each pattern being trained over $800$ cycles using the learning-from-mistake algorithm. If the pattern was not learned within $800$ steps, the training was halted, and a new pattern was trained. 
The results depicted in Figure \ref{fig:ExpAvg} are from different networks utilizing distinct {devices}.  The shaded region shows the average error, which is calculated over the eight different training sessions.

\subsection{IV-curves}

Current-voltage (IV) curves for the {memristive devices} used in training are shown in Figure \ref{fig:IVcurves}. Each IV curve is representative, as variations exist among the {devices} in both experimental and simulated settings. Figure \ref{fig:IVcurves} (a) presents an IV curve collected from the experimental nonlinear memristor, (b) shows a simulated nonlinear memristor, and (c) illustrates a simulated linear memristor.
\begin{figure}[h!] \centering
    \includegraphics[width=1\textwidth]{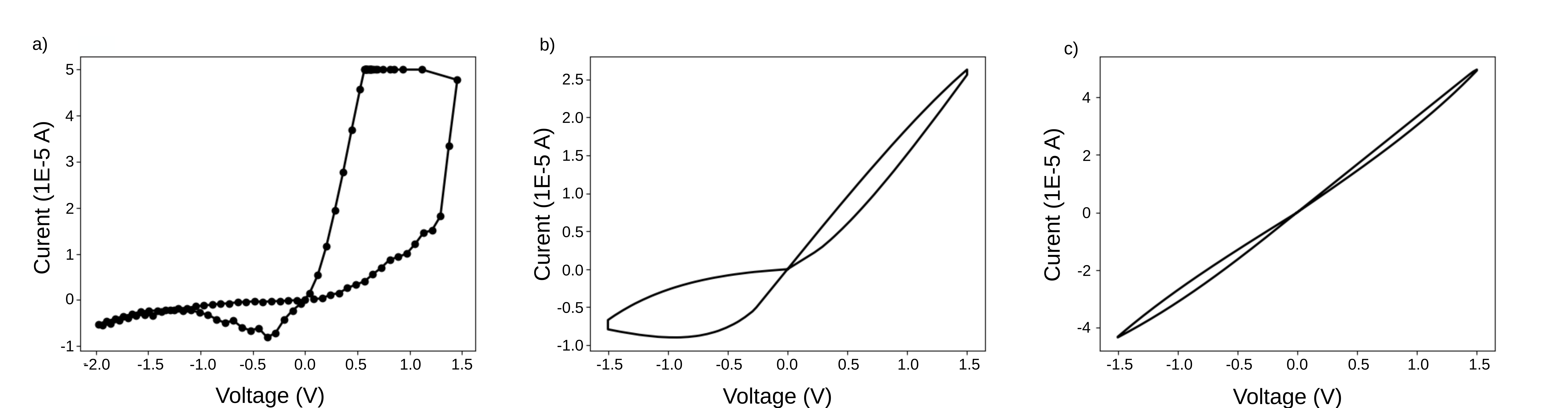}
    \caption{{Examples of experimental, simulated nonlinear, and simulated linear memristor} IV curves are shown in (a), (b), and (c), respectively. The learning parameters and resistance vary across experimental and simulation settings. }
         \label{fig:IVcurves}
\end{figure}

\subsection{Pruned $3$-input $3$-output network}
Results from training the pruned $3$-input $3$-output network are shown in Figure \ref{fig:ErrorPlot_NT3x3}. The pruned $3$-input $3$-output network features broken symmetry in its connectivity, with two {devices} emanating from each node (in the forward direction of {the network}). The network includes pathways from all input nodes to all output nodes, as well as nodes that integrate signals from all pairs of input nodes. The error during training two incompatible patterns is shown in Figure \ref{fig:ErrorPlot_NT3x3} (b). The network is able to routinely learn patterns during training, with {errors} decreasing throughout the training process. As shown in Figure \ref{fig:ErrorPlot_NT3x3} (c), the average memory parameters during training highlight the contrast in adjacent weights that are trained, corresponding to nodes that serve as sources for either high or low resistance edges. 
The contrast in adjacent weights is necessary to satisfy equation \ref{eqn:Inequality1} and \ref{eqn:Inequality2}, below.

\begin{figure}[h!] \centering
    \includegraphics[width=.9\textwidth]{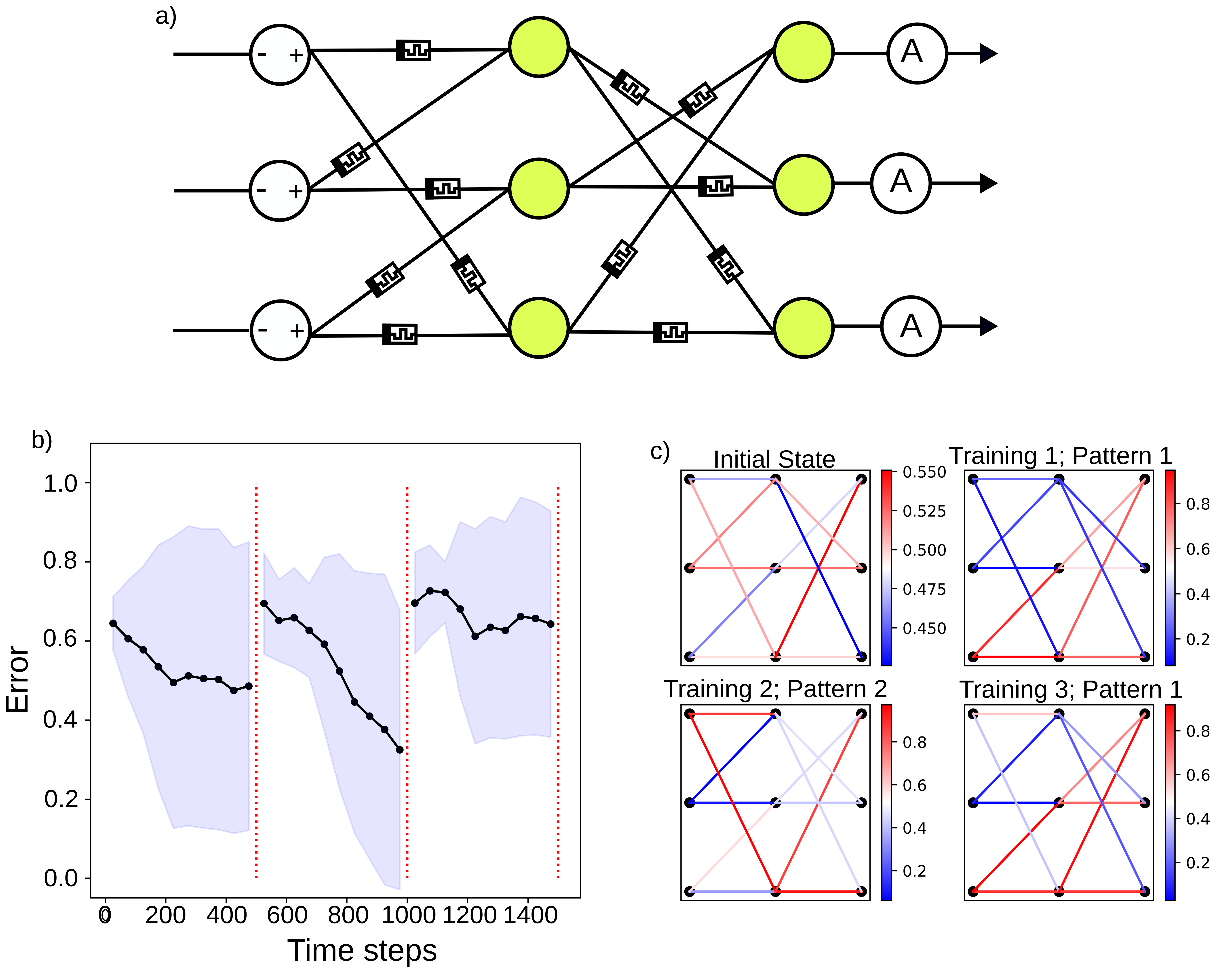}
\caption{ (a) Network schematic of the pruned $3$-input $3$-output network with $3$ nodes in the hidden layer. (b) Average error during training of the pruned $3$-input $3$-output network on two alternating patterns over three training sets, standard deviation shown in shaded region (c). Average values of the memory parameter $x$ in the network at the beginning of training and after training sets. } 
\label{fig:ErrorPlot_NT3x3}
\end{figure}

\subsection{$\Omega_A$ spectral embedding}
As discussed above, under normal operating conditions, $\dot{\vec{x}}$ is the update function of the network during training. This update function is an eigenfunction of the loop projection operator, $\Omega_A$, equation \ref{eqn:LoopEigenfunction}, when $\beta$ is uniform in the network. To assess the controllability of the network, we investigate the independence of the update function by performing a dimensionality reduction of the eigenfunctions of the projection operator, $\Omega_A$. The projection operators $\Omega_A$ correspond to the different circuit connectivities required for distinct corrections during training. The spectral embedding reduces matrices of the eigenfunctions to a two-dimensional subspace. The density of this subspace, along with the overlap of the reduced eigenfunctions, provides insight into how similarly the update functions will operate on the network. For instance, update functions that act similarly on multiple loops will be closer in the lower dimensional subspace. As shown in Figure \ref{fig:SpectralEmbedding}, increasing the size of the network increases the density of the lower dimensional subspace, with values clustered together. The spectral embedding of the pruned networks is sparser compared to fully connected networks with the same number of inputs and outputs. This supports the notion that the pruned networks are more controllable, as the update functions are less similar, allowing corrections during training to better introduce contrast in the network weights.

\begin{figure}[h!] \centering
    \includegraphics[width=.9\textwidth]{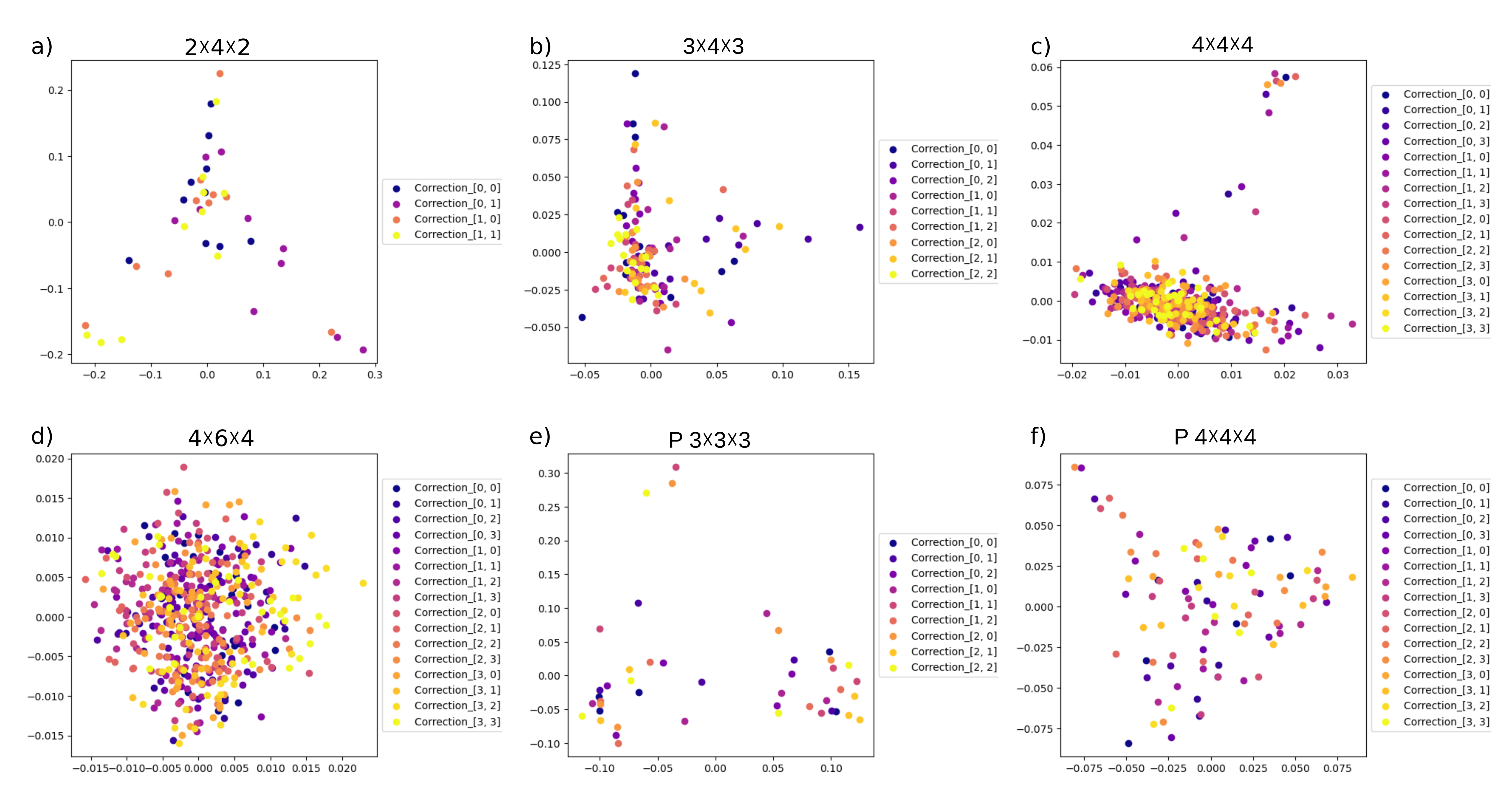}
\caption{Spectral embedding of a matrix of the eigenvalues of the projection operator $\Omega_A$ for corrections during training for networks of different sizes and connectivity. The spectral embedding demonstrates that as the network size increases, there is more overlap and proximity of the dimensionally reduced eigenvalues, indicating that it is challenging to control individual loops in the circuit. In the pruned networks, the spectral embedding is sparse, akin to the $2$-input $2$-output network, demonstrating increased controllability and less overlap of loops in the network.}
     \label{fig:SpectralEmbedding}
\end{figure}

\subsection{Stochastic resources}
{Memristive} networks are rarely uniform {because of sample-to-sample variability in fabrication, and slight variation in the physical parameters lead to differences in each device operation.} These variations include differences in response to applied bias, i.e., the learning rate, and in initial or maximum resistance values. Figure \ref{fig:ErrorsandVariation} shows the average frequency of successfully training a pattern set for networks with and without {memristive device} variation. Simulated {memristive} networks of varying sizes were trained on two alternating patterns over three training eras 20 times. Networks were trained with and without local variation in the learning rate, $\beta$, and in initial resistance values, $x_0$. We observe that local variation acts as a stochastic resource that enables the learning-from-mistakes algorithm to successfully train a pattern. This is because local variation helps drive contrast in the adjacent weights in the network, thus facilitating effective training. Without any variation, in both $\beta$ and $x_0$, the network never learns a pattern. As the network size increases, networks without $x_0$ variation fail to learn patterns, while networks without $\beta$ variation perform comparably to networks with both $\beta$ and $x_0$ variation.

\begin{figure}[h!] \centering
    \includegraphics[width=.5\textwidth]{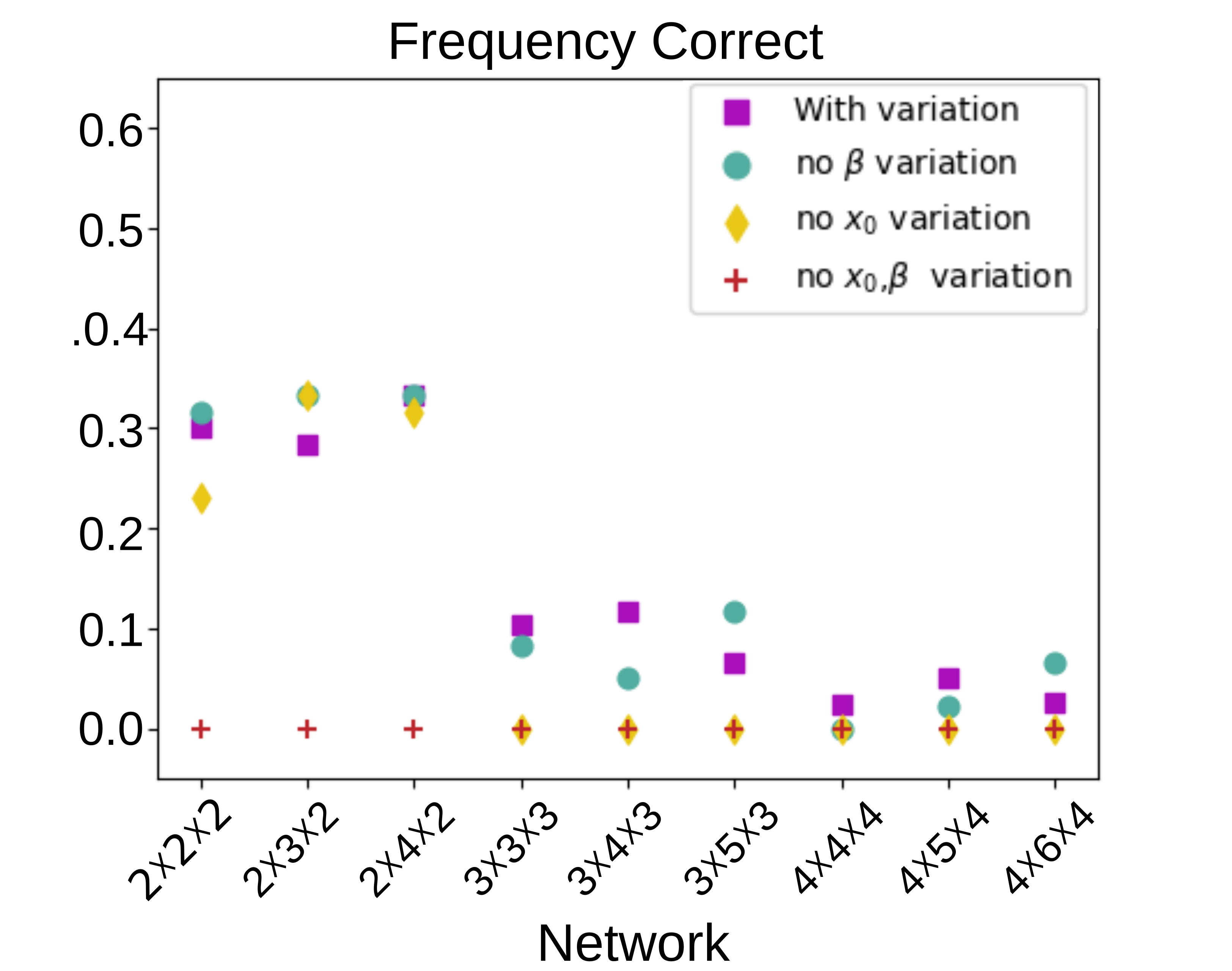}
\caption{Average frequency of successfully training a pattern per training set is shown for different network sizes. Networks with and without random $\beta$ and initial $x$ values ($x_0$) are shown. Without variation in either $\beta$ or $x_0$ the network does not successfully learn a pattern.}
     \label{fig:ErrorsandVariation}
\end{figure}

Examining the inequalities in equations \ref{eqn:Inequality1} and \ref{eqn:Inequality2}, it is apparent that variation in the network's conductivity is necessary to satisfy these inequalities. Neglecting the normalization step, under the ideal conditions where the conductivity of each {device} is only adjusted during the correction step, the parallel paths from the biased input node to the output node experience the same electric potential drop.  Local variation in the conductivity is trained into the network in the learning-from-mistakes algorithm via the update function, $\dot\vec{x}$. For this update function to locally vary, $\vec{x}$ and $\vec{\beta}$ need to vary as seen in equation \ref{eqn:CaravelliEqn}. If the {memristive devices} were initialized with identical resistance, then {those} in a common layer in the parallel paths would evolve under the influences of local $\beta$. Without variation in $\beta$ the conductivities in parallel edges, e.g., $G_{ja}$ and $G_{ia}$, would have similar dynamics. 

This concept is explored explicitly in the simulation above. We find that small networks perform equally well with variation in $\beta$ and initial $x$. As network size increases, having less local variation leads to higher error rates. Without local variation in $x_0$, the $3$-input $3$-output networks fail to learn. The networks never learn a pattern without variation in $\beta$ or initial $x$. There must be inherent local variation in the learning rate so that {each of the devices evolves} in distinct ways within the second layer. This local variation serves as a resource for the update function, driving contrast in the resistivity of the circuit.

Local variation is a crucial resource for networks to learn information. In experiments and simulations, this variation was present in the device learning rate and the initial resistance. The presence of high and low resistance states within the network facilitates the formation of distinct conductive pathways, which are essential for pattern recognition.

\subsection{Correlation analysis of pruned networks}

Figure \ref{fig:MoreCorrelations}, illustrates the positive and negative correlations in the change in effective resistance, shown in red and blue, respectively, as discussed in the main text. Rows (i) and (ii) depict the correlations for fully connected and pruned $3$-input and $3$-output networks, respectively. 
Rows (iii) and (iv) depict the correlations for fully connected and pruned $4$-input networks, respectively.
As in the main text, these are the correlations in the change in $2$-point effective resistance for a given correction throughout the training protocol, where persistent negative and positive correlations are observed throughout training.

Comparing rows (i) and (ii), the fully connected network exhibits stronger correlations overall, whereas the pruned network does not display such large correlations across all the corrections received. Similarly, the fully connected network shows block diagonal correlations, which are absent in the pruned network.
Comparing rows (iii) and (iv), the fully connected network shows more correlations which are present in response to multiple corrections. In the fully connected network, positive correlations appear between effective resistances with a common output node. This feature is absent in the pruned network, where corrections result in opposing correlations. For example, the lower right block diagonal positive correlations in the last three columns of row (v) are disrupted by a checkerboard pattern in the first column. Corrections that produce alternating correlations enable the training protocol to drive contrast in the network weights.

\begin{figure}[h!] \centering
    \includegraphics[width=.9\textwidth]{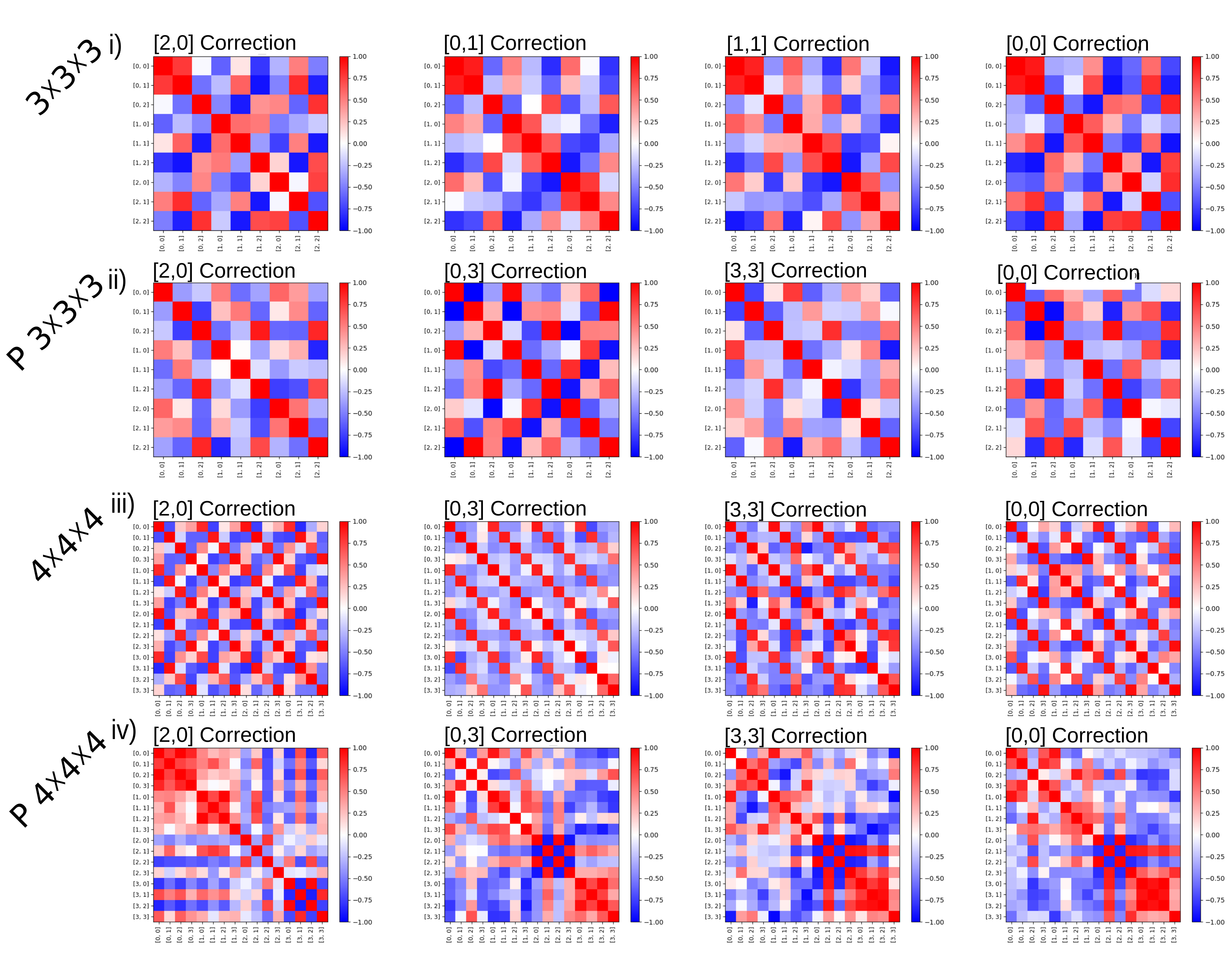}
\caption{Representative examples of correlations throughout training between the $2$-point effective update function in response to a correction, shown in each plot. $2$-point correlations are listed along the $x$ and $y$ axes. Row (i): $3\times 3\times 3$ networks; row (ii): the pruned $3\times 3\times 3$ networks; row (iii): the $4\times 4\times 4$ network; and row (iv): the pruned $4\times 4\times 4$ network. }
     \label{fig:MoreCorrelations}
\end{figure}

\newpage
\clearpage

\section{Supplementary Theory}
\subsection{Capacity}

In the case with winner-take-all outputs, and when inputs are a constant applied bias on individual input nodes, we examine the capacity of the network. 
The network outputs correspond to an $\ArgMax(\vec{i})$, where the maximum current is identified for all output edges. We define a conditional $\ArgMax(i_0 \vert \vec{i})$, which returns $1$ if $i_0=\ArgMax(\vec{i})$ and zero otherwise, assuming distinct values of current on all the edges. This can be generalized to a function to return $1/N$ for $N$ replicas of the maximum value). We express this as
\begin{align}
    \ArgMax(i_i \vert \vec{i}) = \lim_{\frac{1}{k_B T}\rightarrow \infty} \frac{\exp(\frac{1}{k_B T} i_i)}{\sum_j \exp(\frac{1}{k_B T} i_j)}
    \label{eqn:ArgMaxDefine}
\end{align}
where the sum in the denominator is over the outputs of the network. 
Here, we assume that the circuit remains unchanged when biasing or reading distinct mappings, e.g., with a low read bias. Thus, we have a single cycle projection operator $\Omega_A$ and conductivity values $G(x)$ corresponding to the state of the network being measured. 
Without loss of generality, we define our read operation as having all input edges (with voltage generators) connected to all output edges. In general, we can change the network connection for the read operation, e.g., all outputs connected to the biased input. This does not significantly change the results below, where assuming $\Omega_A$ is constant, we can gain a deeper understanding of the network dynamics.

We can rewrite the current one as
\begin{align}
    \vec{i}=-\Roff^{-1} (I-\chi \Omega_A g(x))^{-1}\Omega_A \vec{s} ,
    \label{eqn:CurrentEq}
\end{align}
where $\vec{i}$ is the total current in the network and $\vec{s}$ is the vector of input voltage, corresponding edges with generators. Mappings can be defined by the input edge with nonzero bias, where $s^k$ is the $k$-map, and $s^k$ is a vector of zeros with a non-zero entry on edge $k$.

Substituting equation \ref{eqn:CurrentEq} into equation \ref{eqn:ArgMaxDefine} we have
\begin{align}
    \ArgMax(i_i \vert \vec{i}) &= \lim_{\frac{1}{k_B T}\rightarrow \infty} \frac{\exp(-\frac{1}{k_B T}\Roff^{-1} (I-\chi \Omega_A g(x))^{-1}_{ij}\Omega_A^{jk} \vec{s}_k)}{\sum_o \exp(-\frac{1}{k_B T}\Roff^{-1} (I-\chi \Omega_A g(x))^{-1}_{oj}\Omega_A^{jk} \vec{s}_k)}  ,
\end{align}
here the sum in the denominator is over outputs. We define $\mathcal{P}\equiv (I-\chi \Omega_A x)^{-1}$, and note that we can generalize this to a monotonic function  $g(x)$, with the substitution $x \rightarrow g(x)$. In the limit $\lim_{\frac{1}{k_B T}\rightarrow\infty}$ we have written $ \frac{1}{Roff k_B T}\rightarrow \frac{1}{k_B T}$. This gives us the form used in the main text
\begin{subequations}\begin{align}
    \ArgMax^{s_k}(i_i \vert \vec{i}) &= \lim_{\beta\rightarrow \infty} \frac{\exp(\frac{1}{k_B T}\mathcal{P}_{ij}\Omega_A^{jk} s_k)}{\sum_o \exp(\frac{1}{k_B T}\mathcal{P}_{oj}\Omega_A^{jk} s_k)} .
    \label{eqn:ArgMaxProjection_supp}
\end{align}
\end{subequations}

The mappings and corresponding outputs can be written as a $OUT \times IN$ matrix, for $OUT$-outputs and $IN$-inputs. We have a matrix 
\begin{align}
    A=\begin{pmatrix}
        \ArgMax^{s_0}(i_0\vert\vec{i}), & \dots, & \ArgMax^{s_{IN}}(i_0\vert\vec{i})
        \\ \vdots & \ddots & \vdots 
        \\ \ArgMax^{s_0}(i_{OUT}\vert\vec{i}), & \dots, & \ArgMax^{s_{IN}}(i_{OUT}\vert\vec{i})
    \end{pmatrix}
\end{align}

which takes values of $0$ and $1$. Assuming that each output current for a given mapping is unique in a trained network, each column will have one nonzero value. The number of rules with distinct measured outputs is the capacity of the network. 
The rank of this matrix is the capacity.  The rank depends on the linearly independence of rows and columns of the expectation value of the current. Inserting equation \ref{eqn:ArgMaxProjection_supp} into $A$, it is apparent the rank of $A$ does not relate directly to the rank of $\mathcal{P}\Omega$. 
If each row has a single nonzero value, i.e., the case where each mapping has a distinct output, output, the rank of this matrix is $\text{min}(IN,OUT)$.

We write $A$ as a product 
\begin{subequations}\begin{align}
    A &=\lim_{\frac{1}{k_B T}\rightarrow \infty}
    \begin{pmatrix}
       \frac{\exp(\frac{1}{k_B T}\mathcal{P}_{0j}\Omega_A^{j0}s_0)}{\sum_{i\in {OUT}}\exp(\frac{1}{k_B T}\mathcal{P}_{ij}\Omega_{j0}s_0) } 
       &\dots& 
       \frac{\exp(\frac{1}{k_B T}\mathcal{P}_{0j}\Omega_A^{j {IN}}s_{IN})}{\sum_{i\in R}\exp(\frac{1}{k_B T}\mathcal{P}_{ij}\Omega_{j,IN}s_{IN}) } \\
       \vdots & \ddots &\vdots \\
       \frac{\exp(\frac{1}{k_B T}\mathcal{P}_{{OUT},j}\Omega_A^{j0}s_0)}{\sum_{i\in {OUT}}\exp(\frac{1}{k_B T}\mathcal{P}_{ij}\Omega_{j0}s_0) } 
       &\dots& 
       \frac{\exp(\frac{1}{k_B T}\mathcal{P}_{{OUT},j}\Omega_A^{j{IN}}s_{IN})}{\sum_{i\in {OUT}}\exp(\frac{1}{k_B T}\mathcal{P}_{ij}\Omega_A^{j,{IN}}s_{IN}) } 
    \end{pmatrix}
    \\
    &= BD
\end{align}
\end{subequations}
with $D$ a diagonal matrix with $D_{ii}=\frac{1}{\sum_{i\in {OUT}}\exp(\frac{1}{k_B T}\mathcal{P}_{jk}\Omega_A^{ki}s_i)}$, and 
\begin{align}
    B &= \begin{pmatrix}
       \exp(\frac{1}{k_B T}\mathcal{P}_{0j}\Omega_A^{j0}s_0)
       &\dots& 
       \exp(\frac{1}{k_B T}\mathcal{P}_{0j}\Omega_A^{j,{IN}}s_{IN}) \\
       \vdots & \ddots &\vdots \\
       \exp(\frac{1}{k_B T}\mathcal{P}_{{OUT},j}\Omega_A^{j0}s_0)
       &\dots& 
       \exp(\frac{1}{k_B T}\mathcal{P}_{{OUT},j}\Omega_A^{j,{IN}}s_{IN})
    \end{pmatrix} .
\end{align}
We note that the Frobenius norm is $\vert\vert A\vert\vert=\sqrt{\text{max}({IN},{OUT})}>1$. With positive real current values the exponential is a monotonic function. 
Thus two rows (columns) will be linearly dependent when there is an additive constant $c$ to the output currents in the numerator which equates the currents in both rows (columns). For row, $0$ and $p$, this would be
\begin{equation}
    \begin{pmatrix}\exp{\left(\frac{1}{k_B T}(i^0_0\pm k_B Tc)\right)}&\cdots&\exp{\left(\frac{1}{k_B T}(i^q_0\pm k_B Tc)\right)} \end{pmatrix}=
\begin{pmatrix}\exp{\left(\frac{1}{k_B T}(i^0_p)\right)}&\cdots&\exp{\left(\frac{1}{k_B T} (i^q_p)\right)}\end{pmatrix} .
\end{equation}
This corresponds to a finite difference in the action of the projection operators. for the rows and columns, there exists $a$ and $b$ such that for any row or column $i$, respectively,  
\begin{subequations}
        \begin{align}
          (\mathcal{P}\Omega_A) _{a,i} -(\mathcal{P}\Omega_A) _{b,i} &= k_B T c  ,
   \\
    (\mathcal{P}\Omega_A) _{i,a} -(\mathcal{P}\Omega_A) _{i,b} &= k_B T c .
 \end{align}
\end{subequations}
This difference can be moved out of the exponential as a coefficient of the terms in the matrix,
\begin{subequations}\begin{align}
    \lim_{\frac{1}{k_B T}\rightarrow\infty} c_0 \exp{\left(-\frac{1}{k_B T} \mathcal{P} \Omega_A s\right)}
&= \exp{c}\exp{\left(-\frac{1}{k_B T} \mathcal{P}\Omega_A s\right)}
\\
&= \exp{\left(-\frac{1}{k_B T} (\mathcal{P}\Omega_A s+k_B T c_0\right)} .
\end{align}\end{subequations}
As $\frac{1}{k_B T}\rightarrow \infty$ we have $\vert k_B T c\vert\ll \mathcal{P}\Omega_A$ unless $\vert c \vert\rightarrow \infty$. Linear dependence means there are rows (columns) that can be rewritten as
\begin{align}
\begin{pmatrix}
    \exp{\left(-\frac{1}{k_B T}(i^0_0\pm k_B T c)\right)}&\cdots&\exp{\left(-\frac{1}{k_B T}(i^q_0\pm k_B T c)\right)} \\
    \exp{\left(-\frac{1}{k_B T}(i^0_p\pm k_B T c)\right)}&\cdots&\exp{\left(-\frac{1}{k_B T}(i^q_p\pm k_B T c)\right)}
\end{pmatrix}
\end{align}
This amounts to adding a constant offset to all current outputs on a row (column). In the case of columns, adding a constant current offset to all output edges does not change the measured output of the network. There are linearly independent rows (columns) $a$ and $b$ if
\begin{align}
    \exists a,b \; | \; (\mathcal{P}\Omega_A) _{a,i} -(\mathcal{P}\Omega_A) _{b,i} &= k_B T c, \; \forall i \\
    \exists a,b \; | \; (\mathcal{P}\Omega_A) _{i,a} -(\mathcal{P}\Omega_A) _{i,b} &= k_B T c, \; \forall i .
\end{align}
We can understand this as when two rows (or columns) in $\mathcal{P}\Omega_S$ have a constant offset. This follows from 
\begin{align}
    \sum_j \mathcal{P}_{ij}(\Omega_A^{ja}-\Omega_A^{jb})&= \sum_j (I-\chi\Omega_A x)^{-1}_{ij}(\Omega_A^{ja}-\Omega_A^{jb})
    \nonumber\\ &=k_B T c \; \forall i
    \\
     \sum_j (\mathcal{P}_{aj}-\mathcal{P}_{bj})\Omega_A^{ji}&= \sum_j \left( (I-\chi\Omega_A x)^{-1}_{aj} -(I-\chi\Omega_A x)^{-1}_{bj}\right)\Omega_A^{ji}
    \nonumber\\ &=k_B T c \; \forall i . 
\end{align}
Examining the coefficient $e^{c}$,
as $\frac{1}{k_B T}\rightarrow \infty$ then $\vert k_B T c\vert\rightarrow 0$ unless $\vert c\vert\rightarrow\infty$.
The case with $c\rightarrow -\infty$ amounts to re-scaling the row (column) to zero, corresponding to a trivial linear dependence. The case with $c\rightarrow \infty$ is not permitted as it produces an invalid input to the $\ArgMax$ function. Thus two rows (columns) will be non-trivially linearly dependent if they are identical. In the case of a fully trained network, a circuit with distinct outputs for all distinct inputs, the rank and capacity is $\text{min}(IN,OUT)$. As such we did not investigate networks with different numbers of input and output nodes.

We examine whether the hidden layer can change the rank of the capacity. This would influence if the network can produce distinct outputs for a given number of inputs. We assume that the number of inputs, $IN$, is the same as the number of outputs, $OUT$, labeled as $N$ for now.
The current can be written in terms of the applied bias\cite{barrows_2024},
\begin{equation}
    \vec{i}=-A^t(ARA^t)^{-1}A\vec{v}_\text{source}
\end{equation}
where $R$ is a diagonal matrix with values of the resistance for the edges in the circuit. We assess the linear transformation of this equation by examining the rank of $A^t(ARA^t)^{-1}A$, which is $\text{min}(\text{rank}(A),\text{rank}(ARA^t))$. As $R$ is diagonal, the rank corresponds to the number of cycles in $A$. When the number of cycles is less than the number of output nodes, there is a bottleneck and the circuit cannot possibly generate $N$-distinct outputs for $N$ distinct mappings. Focusing on the two-layer network connecting input and output nodes, with $N>1$, and neglecting other edges including the edges to ground and from voltage generators, we can determine the number of cycles in the network as
\begin{equation}
CYCLES=IN\cdot MID + OUT\cdot MID -( IN + OUT + MID -1)  ,
\end{equation}
where $MID$ is the number of nodes in the hidden layer.  There is a bottleneck when $CYCLES<N$ and 
\begin{subequations}
    \begin{align}
MID &<\frac{3N-1}{2N-1}
\end{align}
\end{subequations}
which for integer values of $N$ converges to $\frac{3}{2}$. We find there is a bottleneck that limits the capacity of a fully connected two-layer circuit when there is only one node in the hidden layer.

\subsection{Unique solutions}

The number of solutions that satisfy a specific mapping includes all configurations of the network in which the effective resistance between a target input and output node is less than the effective resistance between the targeted input node and all other output nodes.

To find the two-point effective resistance we define a partition function $Z$ in terms of node potentials, $\phi$,
\begin{eqnarray}
    Z &=&\int d\phi_0 \cdots d\phi_{n-4} \exp{\left(\frac{-1}{2k_B T}\sum_{i,j}\frac{(\phi_i-\phi_j)^2}{R_{i,j}}\right)} 
    \nonumber \\
    &=& \int d\phi_B \exp{\left( \frac{-1}{2k_B T}\vec{\phi}^t L \vec{\phi}\right)} \nonumber
    \\ &=& \int  d\phi_B \exp{\left(\frac{-1}{2k_B T} \phi_B^t L_{BB} \phi_B+\phi_S^t L_{SS} \phi_S +\phi_S^t L_{SB}\phi_B+\phi_B^t L_{BS} \phi_S\right)}
\end{eqnarray}
where $\phi_B$ is $\vec{\phi}_{0\cdots n-2}$, the vector of electric potentials excluding the two input and output nodes in the effective circuit, e.g., the bulk of the network, and $\phi_S= \vec{\phi}_{n-1\cdots n}$, e.g., the surface of the network. We have written the power in terms of the Laplacian matrix for the circuit, $L$. 
 We can rewrite our Gaussian integral in terms of the bottom $2\times 2$ block values, i.e., the 2-two-point input-output block, of the inverse Laplacian matrix\footnote{Note that this is the result is the same as the bottom corner of a block-wise matrix inversion (Schur inverse),
\begin{eqnarray}
    \begin{pmatrix} A & B \\ C & D \end{pmatrix}^{-1}= 
    \begin{pmatrix}
        A^{-1}+A^{-1}B\left(D-CA^{-1}B\right)^{-1}CA^{-1} & -A^{-1}B\left(D-CA^{-1}B\right)^{-1}
        \\
        -\left(D-CA^{-1}B\right)^{-1} C A^{-1} & \left(D-CA^{-1}B\right)^{-1}
    \end{pmatrix} 
\end{eqnarray}},
 \begin{eqnarray}
     Z\propto \exp{\left( \frac{-\beta}{2}\phi_S^t (L^+_{2\times2})^+\phi_S\right)}.
 \end{eqnarray}
Here $(L^+_{2\times2})^+$ represents the effective circuit, $L^{\text{eff}}$; it is symmetric and, in general, the rows and columns do not sum to $0$. It is not necessarily a singular matrix as the case of a Laplacian matrix. We find the exact effective resistance matrix,  
\begin{eqnarray}
    L^{\text{eff}} &=& B G^\text{eff}B^t \\
    L^+_{2x2} &=& \left(B G^\text{eff}B^t\right)^+
    \\
    G^{\text{eff}+}_{ij} &=& (B^t L^+_{2\times 2} B)_{ij} \nonumber\\
    &=& \tilde{R}^{\text{eff}}_{ij} ,
\end{eqnarray}
where $B$ is the incidence matrix for our effective circuit. As this is a one-edge two-node effective circuit 
\begin{equation}
    \tilde{R}^{\text{eff}} = L^+_{ii}+L^+_{jj}-L^+_{ij}-L^+_{ji} .
\end{equation}

 Focusing on the fully connected two-layer circuit, neglecting connections across the voltage generators and to the ground, we find the effective conductivity is 
\begin{subequations}\begin{align}
    \tilde{G}_{r_i,s_k} &=D_{r_i,s_k}-C_{r_i,j}C_{s_k,j} A^{-1}_{j,j}
    \\
    &=  \sum_j \frac{G_{r_i,j}G_{s_k,j}}{\sum G_j}
\end{align}\end{subequations}
    where $\sum G_{j}$ is the sum of conductivity of all edges connecting to node $j$ in the middle layer.
%

The output currents can be calculated using the input voltage and the effective circuit. The effective conductivity accounts for the parallel paths to a specific output node. The effective output current through node $a$ when voltage is applied to input node $0$ is given by
\begin{align}
    \vec{i}_a = V_0 \tilde{G}_{0a} .
\end{align}

The algorithm's output is determined by identifying the output node with the largest current.  As $V_i$ is constant for individual mappings, the output of the circuit with two inputs $(0,1)$ and two outputs, $(a,b)$, is $[0\rightarrow a]$ and $ [1\rightarrow b]$  when

\begin{subequations}\begin{align}
    \tilde{G}_{0a} &> \tilde{G}_{0b} 
    \\
    \left(\sum_j \frac{G_{0j}G_{ja}}{\sum G_j}\right) &> \left(\sum_j \frac{G_{0j}G_{jb}}{\sum G_j}\right)
    \label{eqn:EffectiveInequalitya}
\end{align}\end{subequations}
and 
\begin{subequations}\begin{align}
    \tilde{G}_{1a} &< \tilde{G}_{1b} 
    \\
    \left(\sum_j \frac{G_{1j}G_{ja}}{\sum G_j}\right) &< \left(\sum_j \frac{G_{1j}G_{jb}}{\sum G_j}\right)
    \label{eqn:EffectiveInequalityb}
\end{align}\end{subequations}    
%

Here this is a passive circuit assuming effective conductivities are all positive. We gain insight into this system by examining a network with just two outputs. The results can be generalized to compare the current in any two outputs, we rewrite the inequalities, equations \ref{eqn:EffectiveInequalitya} and \ref{eqn:EffectiveInequalityb}, as
\begin{align}
    \sum_j \frac{G_{0j}(G_{ja}-G_{jb})}{\sum G_j} &>0
    \\
    \sum_j \frac{G_{1j}(G_{ja}-G_{jb})}{\sum G_j} &<0
\end{align}
The differences in conductivity terms are the differences in conductivity in edges in the second layer emanating from a single hidden node.
As all $G_{ij}$ are positive, in order for both inequalities to be true, at least one of the difference terms is negative and one of the difference terms is positive. These terms are scaled by the coefficients $G_{IN,j}$, the conductivity of edges connected to the biased input nodes. 
This holds for any system where each input maps to a distinct output. In the case of two nodes in a hidden layer, one of the difference terms in the numerator is negative, and the other is positive. We examine this case in more detail. We immediately see that the maximum current depends on the ratio of the conductivity in the input layer and the ratio of the difference in conductivities in the output layers. For the case of two nodes $i,j$ in the hidden layer
\begin{align}
    \frac{G_{0i}}{G_{0j}}\frac{\sum G_j}{\sum G_i} \lessgtr \frac{\vert G_{ja}-G_{jb}\vert}{\vert G_{ia}-G_{ib}\vert}
    \label{eqn:Inequality1}
    \\
    \frac{G_{1i}}{G_{1j}}\frac{\sum G_j}{\sum G_i} \gtrless \frac{\vert G_{ja}-G_{jb}\vert}{\vert G_{ia}-G_{ib}\vert}
    \label{eqn:Inequality2}
\end{align}
 
 For the case of more inputs and outputs, as well as middle layers, these conditions can be generalized to $r(s-1)$ inequalities, for $r$ inputs and $s$ outputs. 

The values of the memristive device conductivities need to be driven apart in order to learn a pattern with distinct outputs. For example, if $\vert G_{ia}-G_{ib}\vert$ is very small, than the ratio $\frac{G_{1i}}{G_{1j}}$ may need to be large. This is more apparent if we investigate the change in conductivities that occurs when learning a new set of patterns, and the conductivities change from $G_{ij}$ to $G^\prime_{ij}$. We have
\begin{align}
\frac{G_{0i}}{G_{0j}}\rho_0 < \frac{\vert G_{ja}-G_{jb}\vert}{\vert G_{ia}-G_{ib}\vert} \rightarrow  \frac{G^{\prime}_{0i}}{G^{\prime}_{0j}}\rho_0^\prime > \frac{\vert G^\prime_{ja}-G^\prime_{jb}\vert}{\vert G^\prime_{ia}-G^\prime_{ib}\vert}
    \\
    \frac{G_{1i}}{G_{1j}}\rho_1 > \frac{\vert G_{ja}-G_{jb}\vert}{\vert G_{ia}-G_{ib}\vert}
    \rightarrow
    \frac{G^{\prime}_{1i}}{G^{\prime}_{1j}}\rho_1^\prime < \frac{\vert G^\prime_{ja}-G^\prime_{jb}\vert}{\vert G^\prime_{ia}-G^\prime_{ib}\vert}
\end{align}
with $\rho_0=\frac{\sum G_j}{\sum G_i}$ and $\rho_1=\rho_1^{-1}$.

We can get a sense that contrasting values of conductivity, e.g., high and low values, are needed to learn a pattern set. This can be seen by noting that there needs to be a finite difference in the conductivity of edges linking a middle node to the output layers, and the finite difference scales the ratio of conductivities in the first layer. In going to a new pattern set these values change, e.g., $\frac{G_{1i}}{G_{1j}}$ increases while $\frac{G^\prime_{1i}}{G^\prime_{1j}}$ decreases.

Examining the simulation results with $\beta$ and initial $\vec{x}_0$ variation, this corresponds to a general pruning mechanism, wherein dominant pathways with high conductivity between input and output edges arise, and pruned pathways (edges with low conductivity) are identified. In learning a pattern set the network learns to prune and reinforce certain paths.

\subsection{Interference}
Here we explore the issue of interference that occurs when learning incompatible mappings. Electrical current can be defined in terms of a matrix $\mathcal{P}\Omega_A$,
\begin{subequations}\begin{align}
    \vec{i} &=-\Roff^{-1}(I-\chi\Omega_A G(x))^{-1}\Omega_A\vec{s}
    \\
    \vec{i} &=\mathcal{P}\Omega_A \vec{s}
\end{align}\end{subequations}
with $\vec{s}$ is a vector with only one non-zero element. We can rescale the applied bias such that the mapping $k$ can be expressed as the delta function, $\vec{s}^k=\delta_{k}$. The output current on edge $i$ depends only on the biased input node. For the mapping $\vec{s}^k$, this relationship is
\begin{align}
    \vec{i}_i=(\mathcal{P}\Omega_A)_{ik} .
\end{align}
The output currents for each mapping in a pattern set $\left\{ s^0,\cdots,s^s\right\}$ can be written as
\begin{align}
    \begin{pmatrix}\vec{i}^t_{s^0} \dots \vec{i}^t_{s^s}
    \end{pmatrix}=\begin{pmatrix}
        \mathcal{P}\Omega_A^{s_0,r_0} & \dots & \mathcal{P}\Omega_A^{s_0,r_r} 
        \\
        \vdots & \ddots & \vdots\\
        \mathcal{P}\Omega_A^{s_s,r_0} & \dots & \mathcal{P}\Omega_A^{s_s,r_r}  .
    \end{pmatrix}
\end{align}
The network's outputs for a specific mapping correspond to the $\ArgMax$ function applied to the output currents for a given input. The network outputs for the entire pattern set are
\begin{subequations}\begin{align}
    \begin{pmatrix}\vec{o}^t_{s^0} \dots \vec{o}^t_{s^s}
    \end{pmatrix} &=\begin{pmatrix}
       \ArgMax( \mathcal{P}\Omega_A^{s_0,r_0}) & \dots & \ArgMax( \mathcal{P}\Omega_A^{s_s,r_0} )
        \\
        \vdots &  & \vdots\\
       \ArgMax( \mathcal{P}\Omega_A^{s_0,r_r}) & \dots & \ArgMax( \mathcal{P}\Omega_A^{s_s,r_r} )
    \end{pmatrix}^t
    \\
    &= A
\end{align}\end{subequations}
where $A$ takes values of $0$ and $1$, and each column has a single nonzero element. For a network with two inputs and two outputs, there are four distinct $A$ matrices:

\begin{align}
    A \in \Biggl\{ \;\; 
    \begin{blockarray}{cc}
        s^0 & s^1 \\
        \begin{block}{(cc)}
            1 & 1 \\
            0 & 0 \\
        \end{block}
    \end{blockarray}
    \;\; , \;\;
    \begin{blockarray}{cc}
        s^0 & s^1 \\
        \begin{block}{(cc)}
            1 & 0 \\
            0 & 1 \\
        \end{block}
    \end{blockarray}
    \;\; , \;\; 
    \begin{blockarray}{cc}
        s^0 & s^1 \\
        \begin{block}{(cc)}
            0 & 0 \\
            1 & 1 \\
        \end{block}
    \end{blockarray}
    \;\; , \;\;
    \begin{blockarray}{cc}
        s^0 & s^1 \\
        \begin{block}{(cc)}
            0 & 1 \\
            1 & 0 \\
        \end{block}
    \end{blockarray}
    \;\; \Biggl\} .
\end{align}
In this representation, the columns correspond to distinct mappings, while the rows correspond to the outputs. 

It is apparent that these four matrices are incompatible, meaning that it is not possible to satisfy two distinct mappings simultaneously. As a result, the system must immediately forget previous mappings to learn a new pattern, leading to catastrophic forgetting.   However, by relaxing the thresholding implemented by $\ArgMax$ such that the outputs are measured with a $\text{Softmax}$ function, we observe a gradual forgetting of previously known mappings, which reduces interference when learning conflicting mappings.

\subsection{Cost function}
Our measurement function corresponds to $\ArgMax$, and can be expressed in terms of a probability 
\begin{align}
    p^a_j=\frac{\exp(\beta \vec{i}^a_j)}{\sum_k \exp(\beta\vec{i}^a_k)} .
\end{align}
Our loss function quantifies the discrepancy between this output probability and the desired mapping, allowing us to apply a correction when this probability disagrees with the desired mapping. We can define the loss function as 
\begin{subequations}\begin{align}
    CE^a_{j}=- p^a_j\ln(\delta^a_{j,g})
    \\CE=-\sum_j p^a_j\ln(\delta^a_{j,g}) .
\end{align}\end{subequations}
Here $\delta^a_{j,g}$ is a delta function for a mapping $a$, with $g$ denoting the desired output. $CE^A_j$ represents the correction applied to an measured output $j$ for mapping $a$. The cost function is binary, there is no error when the $\ArgMax$ output matches the desired output, and a constant error signal is measured whenever the maximum current does not correspond to the desired output, as the correction steps consistently apply the same correction. This cost function is a measure of the cross-entropy loss between our desired and measured current distributions. 

\section{Comparison of Bak-Chialvo Algorithm implementations in the literature}\label{app:algorithmsimp}
Let us first discuss the Chialvo-Bak algorithm as originally envisaged \cite{Chialvo_1999}.

The Chialvo-Bak Learning Machine (CBM) considers that signals propagate through a weighted, directed graph resulting from a Winner-Takes-All (WTA) mechanism by which input signals only traverse edges with the largest weights. Each node's signal propagation follows the direction of outgoing edges, determined solely by the network topology.
As discussed in the original work \cite{Chialvo_1999}, this approach is based on the well-known dynamic competition happening between populations of excitatory and inhibitory neurons each time a given signal is presented to the sensory network inputs. In such a process, only the neurons with stronger connections with the network inputs remain active after a while. The Chialvo-Bak algorithm \cite{Chialvo_1999} simplified this process by directly selecting the nodes with the strongest weights. 
The BC algorithm was designed keeping in mind that the intermediate layer has a large number of neurons, mimicking the so-called divergence in the number of synaptic connections present in biological sensory networks. Thus the BC learning process boils down to choosing the strongest (of an always large pool) weights (i.e., extremal dynamics),  followed by its depression in case of unsuccessful choices.

In the original CBM, signals are binary, indicating node activation status. The memristive version adapts this by using resistors and memristors to represent edge weights. Input signals are introduced via controlled voltage sources, and the output is evaluated by measuring the current through the output nodes.

When propagating a signal, one input is set to signal mode, and the current through the outputs is recorded. The active output is selected based on the WTA criterion, specifically the output with the highest mean squared current over a time period \( T \):

\[ \text{active output} = \arg \max_{n \in \text{Out}} \frac{1}{T} \int_0^T I_n(t)^2 \, dt \]

Learning in the CBM is mistake-driven. The network's performance is compared to a task table, triggering a teaching process when mistakes occur. The teaching mode adjusts memristances to correct errors, ensuring the network learns the correct input-output associations.

Consider a simple network with a generic bulk network with input node 1 and output nodes 2 and 3. If the task requires activation of node 2, but the network incorrectly activates node 3, the teaching process involves a) Opening all inputs and outputs not active during the test b) Setting the correct input to teaching mode c) Applying a constant negative voltage to increase the resistance of the incorrect path.

This adjustment continues until the network consistently activates the correct output. Thus, in WTA Dynamics, Signals follow edges with extreme weights, unlike traditional models where signals are distributed evenly. Moreover, the original CBM uses binary signals, whereas the memristive version uses continuous voltage-based signals. At the end of this procedure, one adjusts memristances based on errors, focusing on correcting specific paths while normalizing the rest.

In this Appendix, we compare the Bak-Chialvo algorithm above as implemented in the attached paper and those implemented in the original \cite{Chialvo_1999}, \cite{Carbajal_2022}, and \cite{Nikiruy_2024}. The algorithm implemented in our manuscript is tailored for a fully memristive neuromorphic hardware setup \textit{without the modified punishment scheme}. 

Let us now refer to \cite{Carbajal_2022}. We provide an explanation taken from \cite{chialvo_bak_learning_1,chialvo_bak_learning_2}, and we refer to those blog posts for a more detailed explanation. We can refer to the picture of our manuscript \ref{fig:KNOWM_Exp}, with nodes $1,2$ as the top and bottom nodes of the input respectively, and nodes $3,4$ as the top out and bottom output top nodes respectively. Let us ignore the trivial case where the resistances are already optimal, and no learning is required. We assume the initial network configuration does not solve the task, necessitating the execution of the teaching steps \(P(1,3)\) or \(P(2,4)\).

Consider \(P(1,3)\) as an example. This teaching step induces currents through all paths from output 3 to input 1. This includes the direct path \(1,3\), which requires increased resistance, as well as other paths necessary for the task, such as \(1,4,2,3\). The figure below illustrates the current directions induced by \(P(1,3)\). In this scenario, the \(1,3\) connection is correctly punished (its resistance will increase), but the \(2,3\) connection is also punished undesirably, and the \(2,4\) connection is rewarded inappropriately. This is a drawback of using a memristive feedforward method because parallel paths are not completely disconnected. We observe that the connection \(1,3\) is punished as intended (its resistance increases), but the \(2,3\) connection is also undesirably punished, while the \(2,4\) connection, which should not be altered, is rewarded.

This example highlights several issues. The WTA dynamics in Chialvo-Bak machines are unsuitable 
for memristive machines because signals propagate through all available paths, not just the one with the lowest resistance. The proposed method can fail in feedforward networks. While it can sometimes succeed, these conditions are nearly equivalent to having the problem already solved. This issue appears to persist in larger networks, as any feedforward network can be reduced to a bipartite graph via equivalent resistances.

Additionally, using only inputs and outputs for teaching proves challenging. It is difficult to punish only the erroneous path without knowledge of the internal topology. As suggested in \cite{Carbajal_2022}, there should be efforts to develop a new teaching method that relies solely on inputs and outputs and investigate network topologies that allow the current method to succeed, which were taken in the present manuscript.

To mitigate undesired effects in simulation, a slight modification to the punishment process has been made in \cite{Carbajal_2022}, e.g.
set to open all edges adjacent to the inputs and outputs that were not active in the test. This is the same scheme adopted in \cite{Nikiruy_2024}, which was implemented experimentally using a crossbar array, with the modified punishment scheme implemented directly in hardware.
This change ensures that current flows only through the undesired connection, improving the effectiveness of the punishment process. However, some adverse initial conditions remain, particularly when the initial resistances of incorrect connections are close to their maximum values. In such cases, the punishment may not sufficiently increase the resistance to solving the task. This problem can be addressed by "priming the network," which involves propagating signals with average values across the network to reduce initial resistances.

\end{document}